\documentclass[sigconf]{acmart}

\AtBeginDocument{%
  \providecommand\BibTeX{{%
    \normalfont B\kern-0.5em{\scshape i\kern-0.25em b}\kern-0.8em\TeX}}}
\usepackage{textcomp}
\usepackage{multirow}
\usepackage{enumitem}
\usepackage{colortbl}

\newcommand{\ie}{\emph{i.e., }}

\definecolor{ours_blue}{RGB}{236, 244, 255}

\copyrightyear{2026}
\acmYear{2026}
\setcopyright{cc}
\setcctype{by} 
\acmConference[CIKM '26]
  {Proceedings of the 35th ACM International Conference on Information and Knowledge Management}
  {November 7--11, 2026}
  {Rome, Italy.}
\acmBooktitle{Proceedings of the 35th ACM International Conference on Information and Knowledge Management (CIKM '26), November 7--11, 2026, Rome, Italy}
\acmISBN{979-8-4007-2539-5/2026/11}
\acmDOI{10.1145/3799682.3841103}

\begin{document}

\title{Latent Residual Reasoning for Ranking in Generative Recommendation}

\author{Haihan Shi}
\affiliation{%
  \institution{University of Science and Technology of China}
  \city{Hefei}
  \country{China}}
\email{shh924@mail.ustc.edu.cn}

\author{Ruidong Han}
\affiliation{%
  \institution{Meituan}
  \city{Beijing}
  \country{China}}
\email{hanruidong@meituan.com}

\author{Wenjie Wang}
\affiliation{%
  \institution{University of Science and Technology of China}
  \city{Hefei}
  \country{China}}
\email{wenjiewang96@gmail.com}

\author{Shangyu Chen}
\affiliation{%
  \institution{Meituan}
  \city{Beijing}
  \country{China}}
\email{chenshangyu03@meituan.com}

\author{Sunhao Dai}
\authornote{Corresponding authors.}
\affiliation{%
  \department{Gaoling School of Artificial Intelligence}
  \institution{Renmin University of China}
  \city{Beijing}
  \country{China}}
\email{sunhaodai@ruc.edu.cn}

\author{Yongqi Li}
\affiliation{%
  \institution{The Hong Kong Polytechnic University}
  \city{Hong Kong SAR}
  \country{China}}
\email{liyongqi0@gmail.com}

\author{Yangyang Li}
\affiliation{%
  \institution{Academy of Cyber}
  \city{Beijing}
  \country{China}}
\email{liyangyang@live.com}

\author{Fei Jiang}
\authornotemark[1]
\affiliation{%
  \institution{Meituan}
  \city{Beijing}
  \country{China}}
\email{jiangfei05@meituan.com}

\author{Wei Lin}
\affiliation{%
  \institution{Meituan}
  \city{Beijing}
  \country{China}}
\email{linwei31@meituan.com}

\author{Fuli Feng}
\affiliation{%
  \institution{University of Science and Technology of China}
  \city{Hefei}
  \country{China}}
\email{fulifeng93@gmail.com}

\author{Xiangnan He}
\affiliation{%
  \institution{University of Science and Technology of China}
  \city{Hefei}
  \country{China}}
\email{xiangnanhe@gmail.com}
\renewcommand{\shortauthors}{Haihan Shi et al.}

\begin{abstract}
Generative recommendation (GR) models for ranking typically adopt a fast-thinking direct prediction paradigm with strong practical efficiency. However, ranking requires fine-grained scoring across multiple candidate items, while existing GR models still mainly rely on one-shot prediction. This raises the question whether additional inference-time computation can further improve ranking quality. Recently, inspired by Chain-of-Thought (CoT) reasoning in large language models (LLMs), some work has begun to introduce reasoning into recommendation systems. Some of these methods rely on LLM reasoning frameworks to perform explicit reasoning in textual space, but such designs are not suitable for more widely deployed non-textual recommendation models. A more relevant direction is to introduce latent reasoning into ID-based recommendation models. Although recent work has extended latent reasoning to industrial ranking, how successive reasoning steps should improve fine-grained candidate scores remains less explored. To address this problem, we propose \textbf{Refiner}, a latent residual reasoning framework with parameter-efficient fine-tuning for generative ranking. Our approach introduces three key components: (1) a residual reasoning mechanism that progressively refines ranking scores through cumulative additive corrections, (2) history queries that provide condensed user behavior representations to enhance candidate-specific reasoning, and (3) a parameter-efficient fine-tuning strategy that leverages LoRA modules to adapt pre-trained models for latent reasoning. Extensive experiments on two public benchmarks, an offline industrial dataset, and an online A/B test demonstrate that Refiner consistently improves over direct prediction, validating its effectiveness and practical utility.
\end{abstract}


\maketitle

\section{Introduction}\label{sec:introduction}
In the era of information explosion, recommender systems have become fundamental components of online web platforms~\cite{singer2022sequential,dai2024modeling,lei2021semi,zhai2024actions,han2025mtgr,zhang2024saqrec,song2026mtfm} by providing users with personalized item suggestions. Among various recommendation architectures, generative recommendation (GR)~\cite{zhai2024actions,han2025mtgr,rajput2023recommender,huang2025towards,deng2025onerec,wang2024learnable,guo2025onesug,wang2025mtgrboost,song2026mtfm} has garnered significant interest from both academia and industry.

For the ranking task in recommendation, some GR methods~\cite{han2025mtgr,zhai2024actions,song2026mtfm} employ a discriminative loss for modeling, utilizing a streamlined Transformer~\cite{vaswani2017attention} architecture. In the setting considered in this work, user context and explicit candidate items are organized into a shared token sequence, while a candidate mask enables parallel candidate scoring without information leakage across candidates. This approach improves system utilization during both training and inference and has shown strong effectiveness in large-scale recommendation systems. However, these methods predominantly adopt a fast-thinking direct prediction paradigm, where the final result is obtained through a single inference step. Since ranking requires fine-grained scoring over candidate items, a key question is whether additional inference-time computation can further refine these scores beyond one-shot prediction.
\par Inspired by the success of the Chain-of-Thought~\cite{wei2022chain} (CoT) mechanism in Large Language Models~\cite{achiam2023gpt,liu2024deepseek} (LLMs), recent work has begun exploring the integration of reasoning into recommendation models by introducing multiple intermediate reasoning steps before generating final recommendations. Some of these methods~\cite{fang2025reason4rec,bismay2024reasoningrec,gu2025r} employ LLMs as recommendation models and introduce mechanisms analogous to explicit CoT in LLM, allowing the recommender to perform reasoning in textual space before generating final recommendations. However, such methods are highly dependent on the reasoning techniques inherent in LLMs~\cite{kojima2022large}, which come with a significant computational overhead~\cite{su2025token}. In addition, LLMs have not completely replaced more widely deployed non-textual recommendation models, limiting the applicability of these approaches in widely deployed ranking scenarios. Therefore, a more practical direction is to introduce latent reasoning~\cite{hao2024training}, \ie multi-step computation in continuous latent space, into widely used Transformer-based recommendation models, including ID-based recommendation settings. Existing attempts along this direction target autoregressive semantic-ID generation~\cite{lin2026care} or refine user representations for next-item recommendation~\cite{tang2025think,shan2026calr,liu2025lares,tang2026parallel,yang2026beyond} with Transformer-based recommenders~\cite{kang2018self,hou2022towards}. These task interfaces differ from ranking with explicit candidates, where user context and candidate items are jointly encoded to directly produce candidate scores. The OnePiece framework~\cite{dai2026onepiece} further demonstrates the feasibility of latent reasoning in industrial ranking through block-wise candidate-representation refinement and progressive multi-task supervision.

Despite this progress, designing latent reasoning for fine-grained candidate scoring raises three challenges. First, latent reasoning must be aligned with the learning objective of candidate ranking. Whereas semantic-ID generation and next-item recommendation optimize different objectives, ranking requires fine-grained scores for explicit candidates. Therefore, successive reasoning steps should be formulated and supervised to progressively improve candidate scores under the ranking objective. Second, in the ranking process, multiple candidate items are scored simultaneously. An intuitive solution is to independently assign one or more reasoning tokens to each candidate item, which we refer to as the candidate-specific reasoning process. However, such a design may fail to effectively leverage global user information, which plays a central role in existing retrieval-oriented latent reasoning. Third, existing latent reasoning frameworks predominantly rely on end-to-end training, which requires replacing or retraining well-established online recommenders and thus limits practical applicability. Consequently, another critical challenge is how to enable latent reasoning through parameter-efficient adaptation with minimal disruption.

To address the above challenges, we propose \textbf{Refiner}, a latent \underline{\textbf{Re}}sidual reasoning framework with parameter-efficient \underline{\textbf{fine}}tuning for \underline{\textbf{r}}anking in generative recommendation. To align latent reasoning with the ranking objective, we design a \textit{residual latent reasoning} mechanism that performs additive refinement in score space. Instead of predicting full scores from scratch at each step, Refiner estimates corrections to the preceding predictions and accumulates them for final ranking. To complement the candidate-specific reasoning process, we further introduce \textit{history queries} to provide condensed user behavior representations that supply a global view of user interests during candidate-specific reasoning. Finally, we develop a parameter-efficient fine-tuning framework by freezing the pre-trained base model and introducing lightweight LoRA~\cite{hu2022lora} modules together with a dedicated reasoning prediction layer, which together enable latent reasoning with minimal disruption to existing recommender systems.
We implement Refiner on two ranking models, including the deployed industrial GR model MTGR~\cite{han2025mtgr}. Extensive experiments on two public benchmarks validate its general effectiveness. Furthermore, Refiner demonstrates consistent improvements on a large-scale offline industrial dataset and in an online A/B test within an industrial setting. We also observe notable gains in scenarios involving cold-start items and long-sequence users with complex behavioral patterns. The contributions of this work are summarized as follows:
\begin{itemize}[leftmargin=*]

\item We investigate latent reasoning for fine-grained score refinement in generative ranking beyond direct one-shot prediction.

\item We propose \textbf{Refiner}, a residual latent reasoning framework for generative ranking that combines multi-step residual score refinement, candidate-specific reasoning with history queries, and parameter-efficient fine-tuning.

\item Extensive experiments on two public datasets, a large offline industrial dataset, and an online A/B test demonstrate consistent improvements, validating the effectiveness and practical utility of Refiner in real-world recommendation scenarios.
\end{itemize}

\section{Preliminary} \label{sec:pre}
In the ranking stage of recommender systems, recommendation models take user-side and candidate item information as input to produce interaction scores for each candidate item across multiple downstream tasks such as Click-Through Rate (CTR) and Click-Through Conversion Rate (CTCVR) prediction. Unlike traditional approaches~\cite{lian2018xdeepfm,zhou2018deep,guo2017deepfm}, GR ranking models formulate the ranking task as a sequential transduction task~\cite{zhai2024actions,han2025mtgr} that organizes input data as a token sequence and processes information in a temporal manner, while simultaneously scoring multiple candidate items within a single forward pass. This design improves gradient estimation robustness during training and enables efficient multi-candidate scoring during inference.

Formally, for a user with profile features and behavior sequences, along with $K$ candidate items, the input data is organized as:
\begin{itemize}[leftmargin=*]
    \item User profile features: $U = [U^1, \ldots, U^{N_U}]$, where $N_U$ denotes the number of user profile features.
    \item User historical interaction sequence: $\overrightarrow{S} = [S^1, \ldots, S^{N_S}]$, where $N_S$ represents the sequence length.
    \item Candidate item features: $I = [I_1, \ldots, I_K]$, where $I_i$ represents the $i$-th candidate and $K$ is the number of candidates.
\end{itemize}

The \textit{embedding module} $\mathcal{E}$ projects all categorical and numerical features into dense vector representations through an embedding process:
\[
\mathbf{E}_U = \mathcal{E}(U), \quad \mathbf{E}_S = \mathcal{E}(\overrightarrow{S}), \quad \mathbf{E}_I = \mathcal{E}(I).
\]
These embedded features are then unified into a single token sequence:
\[
\mathbf{X} = [\mathbf{E}_U, \mathbf{E}_S, \mathbf{E}_I] \in \mathbb{R}^{L \times d},
\]
where $L$ is the total sequence length and $d$ is the embedding dimension.

To enable simultaneous scoring of multiple candidates while preventing information leakage between them, generative ranking employs a \textit{candidate mask} $\mathbf{M} \in \{0,1\}^{L \times L}$ in the attention mechanism. This mask ensures that all candidate items can attend to all user-side information while prohibiting attention between different candidate items. The masked token sequence is then processed by \textit{Transformer blocks} $\mathcal{T}$:
\[
\mathbf{H} = \mathcal{T}(\mathbf{X}, \mathbf{M}) \in \mathbb{R}^{L \times d}.
\]

Finally, the model selects the output representations corresponding to candidate item positions and passes them through a \textit{prediction module} $\mathcal{P}$:
\[
\mathbf{S} = \mathcal{P}(\mathbf{H}[-K:]) \in \mathbb{R}^{K \times T},
\]
where $\mathbf{S}$ contains the task-specific prediction scores for all $K$ candidates across $T$ different tasks. The base model is optimized with Binary Cross-Entropy (BCE) supervision over the task labels:
\[
\mathcal{L}_{\text{Base}} = -\frac{1}{K \cdot T}\sum_{i=1}^K \sum_{j=1}^T \left[ \mathbf{L}_{i,j} \cdot \log(\mathbf{S}_{i,j}) + (1 - \mathbf{L}_{i,j}) \cdot \log(1 - \mathbf{S}_{i,j}) \right],
\]
where $\mathbf{L} \in \{0,1\}^{K \times T}$ denotes the corresponding ground truth labels.

Based on this formulation, latent reasoning introduces additional computation in continuous hidden space without generating textual traces. Refiner uses the resulting states to produce successive corrections to the candidate scores, as described in the following section.

\section{Method}\label{sec:method}

In this section, we propose \textbf{Refiner}, a residual latent reasoning framework with parameter-efficient fine-tuning for generative ranking recommender systems. The framework consists of three core components: (1) Input Construction with reasoning tokens and history queries, (2) Parameter-efficient Fine-tuning Framework, and (3) Residual Reasoning for progressive score refinement and optimization.

\subsection{Input Construction} \label{sec:ic}

To enable latent reasoning in generative ranking, we extend the input sequence by introducing reasoning tokens and history queries. The key idea is to support candidate-specific latent computation while preserving access to shared user context, which is important for ranking multiple candidates in parallel. Accordingly, reasoning tokens are used to carry candidate-specific intermediate states, while history queries provide a compact representation of global user behavior during the reasoning process.

The initial input sequence is constructed as:
\[
\mathbf{X}_0 = [\mathbf{E}_U, \mathbf{E}_S, \mathbf{E}_I] \in \mathbb{R}^{L \times d},
\]
where $L$ is the sequence length, $d$ is the embedding dimension, and $\mathbf{M}_0$ is the original candidate mask that prevents information leakage between candidate items.

\textbf{Reasoning Tokens.} For each reasoning step $r \geq 1$, we append $K$ reasoning tokens $\mathbf{E}_{R_r}$ to the sequence, each corresponding to a candidate item. These tokens are generated auto-regressively from the previous step's output. Specifically, for step $r$, the reasoning tokens $\mathbf{E}_{R_r}$ are obtained from the last $K$ positions of the output hidden states $\mathbf{H}_{r-1}$ of the previous step, where $\mathbf{H}_{r-1}$ is computed as:
\[
\mathbf{H}_{r-1} = \mathcal{T}(\mathbf{X}_{r-1}, \mathbf{M}_{r-1}),
\]
\[
\mathbf{E}_{R_r} = \mathbf{H}_{r-1}[-K:].
\]
This iterative process enables candidate-specific latent reasoning conditioned on user-side information and previous reasoning states.

\textbf{History Queries.} To enrich the reasoning process with condensed user context, we introduce history queries $\mathbf{E}_{HQ}$ from the user historical sequence. We first process the initial sequence $\mathbf{X}_0$ with the Transformer $\mathcal{T}$ to obtain output hidden states:
\[
\mathbf{H}_0 = \mathcal{T}(\mathbf{X}_0, \mathbf{M}_0).
\]
Then, we extract the output representations from the last $N_{\text{HQ}}$ positions of the user historical sequence segment (i.e., the part corresponding to $\mathbf{E}_S$) to form the history queries:
\[
\mathbf{E}_{HQ} = \mathbf{H}_0[\text{start}_{\text{HQ}} : \text{end}_{\text{HQ}}],
\]
where $\text{start}_{\text{HQ}}$ and $\text{end}_{\text{HQ}}$ denote the start and end indices of the last $N_{\text{HQ}}$ tokens in $\mathbf{E}_S$. This fixed set of history queries captures condensed user behavior information and is reused across all reasoning steps.

The input sequence for the $r$-th reasoning step ($r \geq 1$) is constructed by concatenating the initial features, history queries, and all reasoning tokens up to step $r$:
\[
\mathbf{X}_{r} = [\mathbf{E}_U, \mathbf{E}_S, \mathbf{E}_I, \mathbf{E}_{HQ}, \mathbf{E}_{R_1}, \ldots, \mathbf{E}_{R_r}].
\]

For the mask configuration at step $r$, we extend the original candidate mask $\mathbf{M}_0$ to form $\mathbf{M}_r$. Specifically, the mask ensures that:
\begin{itemize} [leftmargin=*]
    \item Each reasoning token at step $r$ can attend to all user-side information ($\mathbf{E}_U$, $\mathbf{E}_S$, $\mathbf{E}_{HQ}$), its corresponding candidate item, and all previous reasoning tokens from steps $1$ to $r-1$ for the same candidate to prevent information leakage.
    \item History queries follow the same attention pattern as $\mathbf{E}_S$, with a causal mask applied within the $\mathbf{E}_{HQ}$ segment to maintain temporal consistency.
\end{itemize}
\begin{figure}[H]
\setlength{\abovecaptionskip}{0cm}
\setlength{\belowcaptionskip}{-0.5cm}

\centering
\includegraphics[scale=0.415]{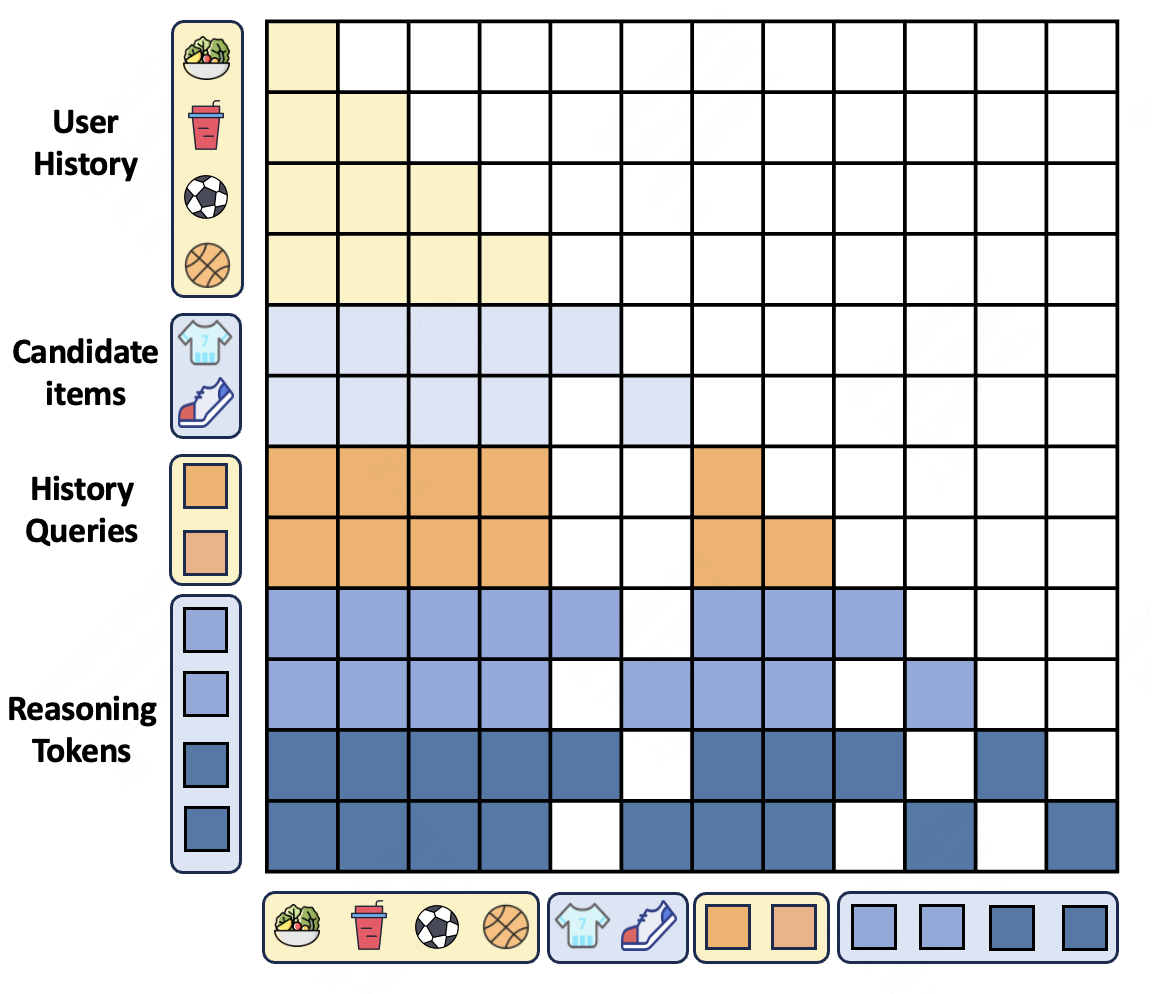}
\caption{The extended candidate mask in Refiner to prevent information leakage during the reasoning process.  History Queries are designed to attend to all User History information while maintaining a causal mask within the query segment itself. The newly introduced Reasoning Tokens, however, are constrained to only access their corresponding Candidate Item, preceding Reasoning Tokens from earlier steps, and all user-side information including User History and History Queries.}
\label{fig:mask}
\end{figure}
An example of the reasoning mask is presented in Figure~\ref{fig:mask}.  

\subsection{Parameter-efficient Fine-tuning Framework}

Since deployed ranking models are often well-established and costly to fully retrain or replace, Refiner introduces latent reasoning through lightweight adaptation rather than end-to-end replacement. To this end, we propose a fine-tuning framework that freezes most parameters of a pre-trained base model and introduces only a small set of trainable components. Let the pre-trained base model consist of an embedding module $\mathcal{E}_{\text{pretrain}}$, Transformer blocks $\mathcal{T}_{\text{pretrain}}$, and a prediction layer $\mathcal{P}_{\text{pretrain}}$. During fine-tuning, we freeze these components to preserve the original ranking behavior of the base model and avoid unnecessary interference.

We then enhance the model with two additions:
\begin{itemize} [leftmargin=*]
    \item \textbf{LoRA-Enhanced Transformer:} We integrate Low-Rank Adaptation~\cite{hu2022lora} (LoRA) into the Transformer blocks, creating an enhanced version $\mathcal{T}_{\text{reasoning}} = \mathcal{T}_{\text{pretrain}} + \text{LoRA}(\mathcal{T}_{\text{pretrain}})$. This introduces a small number of trainable parameters while keeping the pre-trained backbone frozen.
    \item \textbf{Reasoning Prediction Layer:} We add a dedicated prediction layer $\mathcal{P}_{\text{reasoning}}$ for the reasoning steps, which is trained to output residual scores.
\end{itemize}

The reasoning process involves a roll-out mechanism for generating reasoning tokens and history queries:
\begin{itemize} [leftmargin=*]
    \item For the first reasoning step ($r=1$), we use the frozen $\mathcal{T}_{\text{pretrain}}$ to process the initial input $\mathbf{X}_0$ and obtain $\mathbf{H}_0 = \mathcal{T}_{\text{pretrain}}(\mathbf{X}_0, \mathbf{M}_0)$. We then extract history queries $\mathbf{E}_{HQ}$ from the user historical segment of $\mathbf{H}_0$ (as described in Section~\ref{sec:ic}) and first-step reasoning tokens $\mathbf{E}_{R_1}$ from the last $K$ positions of $\mathbf{H}_0$.
    \item For subsequent reasoning steps ($r \geq 2$), we use the enhanced $\mathcal{T}_{\text{reasoning}}$ to process the previous input $\mathbf{X}_{r-1}$ and obtain $\mathbf{H}_{r-1} = \mathcal{T}_{\text{reasoning}}(\mathbf{X}_{r-1}, \mathbf{M}_{r-1})$. We then extract the next reasoning tokens $\mathbf{E}_{R_r}$ from the last $K$ positions of $\mathbf{H}_{r-1}$.
\end{itemize}

This design enables latent reasoning through parameter-efficient adaptation, while preserving the deployed base ranker and minimizing disruption to existing recommender systems.
\begin{figure*}[!t]
\setlength{\abovecaptionskip}{0cm}
\setlength{\belowcaptionskip}{-0.2cm}
\centering
\includegraphics[scale=0.5]{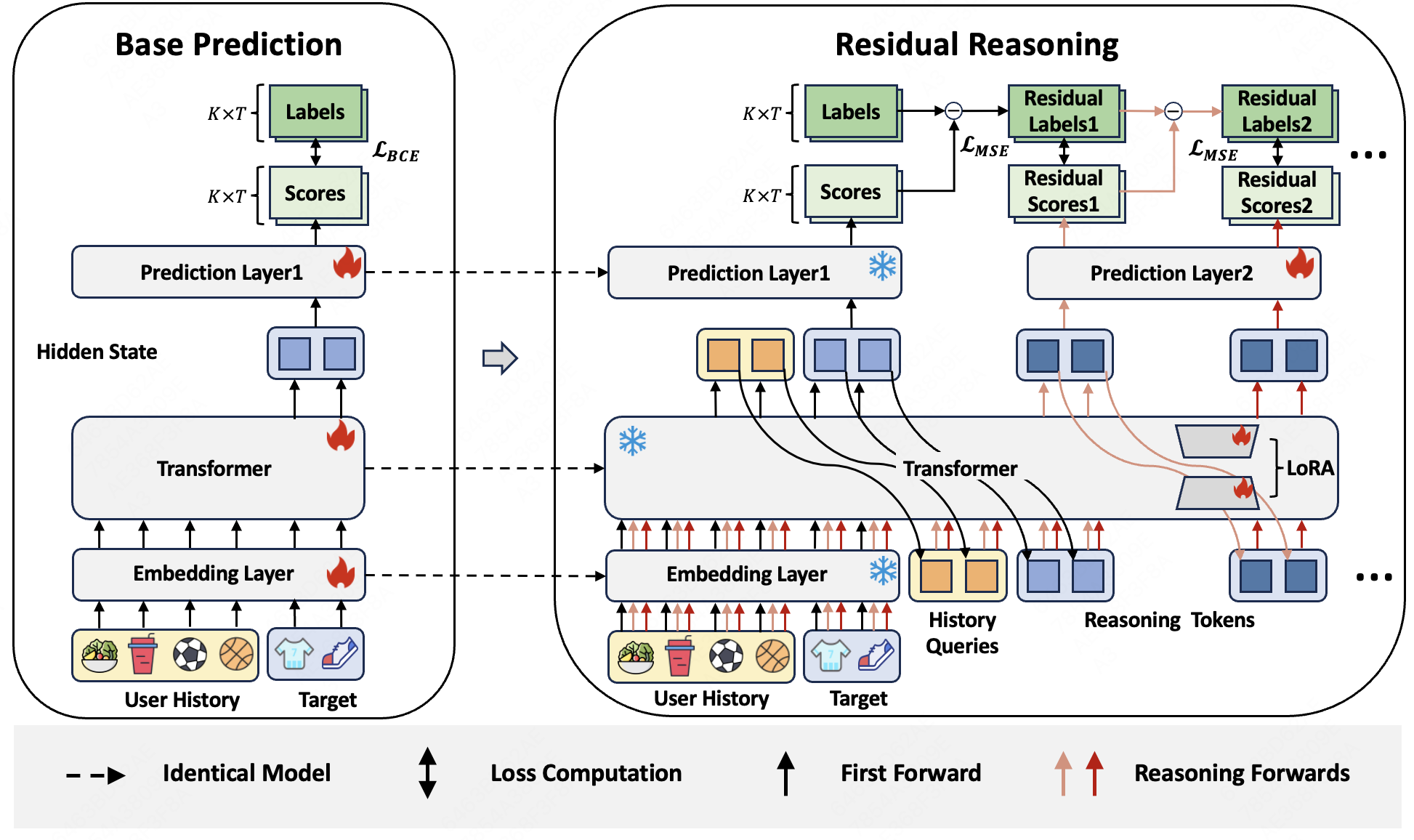}
\caption{Overview of the proposed Refiner framework, which consists of three stages: (a) Input construction with reasoning tokens and history queries, (b) Parameter-efficient fine-tuning with a frozen base model and added LoRA/prediction modules, and (c) Residual reasoning that progressively refines prediction scores through auto-regressive steps.}
\label{fig:main_method}
\end{figure*}
\subsection{Residual Reasoning}

The residual reasoning component aims to progressively refine prediction scores through multiple reasoning steps. In ranking tasks, small but precise score adjustments can be important for final candidate ordering. Therefore, instead of re-predicting the full scores from scratch at each reasoning step, Refiner models latent reasoning as an additive score correction process over the pre-trained model output.

Let $\mathbf{S}_0$ denote the initial score matrix from the pre-trained model, computed as:
\[
\mathbf{S}_0 = \mathcal{P}_{\text{pretrain}}(\mathcal{T}_{\text{pretrain}}(\mathbf{X}_0, \mathbf{M}_0)[-K:]),
\]
where $\mathcal{T}_{\text{pretrain}}(\mathbf{X}_0, \mathbf{M}_0)[-K:]$ refers to the output hidden states at the candidate positions.

For each reasoning step $r \geq 1$, we compute a refinement score matrix $\mathbf{S}_r$ as follows:
\begin{itemize} [leftmargin=*]
    \item The input sequence $\mathbf{X}_r$ (constructed as in Section~\ref{sec:ic}) is processed by the enhanced Transformer $\mathcal{T}_{\text{reasoning}}$ to obtain hidden states:
    \[
    \mathbf{H}_r = \mathcal{T}_{\text{reasoning}}(\mathbf{X}_r, \mathbf{M}_r).
    \]
    \item The output representations at the candidate positions (i.e., the last $K$ positions of $\mathbf{H}_r$) are passed through the reasoning prediction layer $\mathcal{P}_{\text{reasoning}}$ to yield $\mathbf{S}_r$:
    \[
    \mathbf{S}_r = \mathcal{P}_{\text{reasoning}}(\mathbf{H}_r[-K:]).
    \]
\end{itemize}

The cumulative refined score up to step $r$ is defined as the sum of the initial score and all refinement scores from step $1$ to $r$:
\[
\mathbf{S}_{\text{cum},r} = \mathbf{S}_0 + \sum_{i=1}^{r} \mathbf{S}_i.
\]
Each reasoning step is trained to reduce the remaining prediction error through this cumulative score refinement process. Since the reasoning head predicts continuous additive score corrections, we optimize the refinement objective using the mean squared error (MSE) between $\mathbf{S}_{\text{cum},r}$ and the ground truth label matrix $\mathbf{L} \in \{0,1\}^{K \times T}$:
\[
\mathcal{L}_{\text{MSE}_r} = \frac{1}{K \cdot T}\sum_{i=1}^K \sum_{j=1}^T (\mathbf{S}_{\text{cum},r,i,j} - \mathbf{L}_{i,j})^2.
\]
The total loss for all reasoning steps (from $r=1$ to $R_{\text{steps}}$, where $R_{\text{steps}}$ is the total number of reasoning steps) is the sum of the individual MSE losses:
\[
\mathcal{L}_{\text{reasoning}} = \sum_{r=1}^{R_{\text{steps}}} \mathcal{L}_{\text{MSE}_r}.
\]
This residual optimization encourages each reasoning step to incrementally reduce the remaining prediction error, enabling progressive score refinement for ranking tasks.

The overall training framework of Refiner is illustrated in Figure~\ref{fig:main_method} with $N_{\text{HQ}}=2$, $R_{\text{steps}}=2$ and $K=2$. During the residual reasoning phase, Refiner enhances the pre-trained model by incorporating LoRA modules and an additional prediction head $\mathcal{P}_{\text{reasoning}}$ for latent reasoning fine-tuning. The black arrows represent the direct prediction pathway, where Refiner utilizes the fully frozen base model to obtain initial scores requiring refinement, along with the representations used to construct $\mathbf{E}_{R_1}$ and $\mathbf{E}_{HQ}$. During reasoning steps 1 and 2, Refiner auto-regressively derives subsequent reasoning tokens through $\mathcal{T}_{\text{reasoning}}$ and performs residual score correction.

\section{Experiments}
We conduct extensive experiments and analyses on public and industrial datasets to answer the following research questions:
\begin{itemize}[leftmargin=0.5cm]
\item \textbf{RQ1:} How does Refiner perform compared to strong baselines, including latent reasoning methods?
\item \textbf{RQ2:} How do the different components of Refiner affect the performance?
\item \textbf{RQ3:} How robust is Refiner across different user and item groups?
\item \textbf{RQ4:} How do key hyperparameters (\ie the number of history queries and total reasoning steps) affect the performance and inference-time overhead of Refiner?
\item \textbf{RQ5:} How effective and practical is Refiner in real-world industrial scenarios?
\end{itemize}

\subsection{Experimental Setting}\label{sec:exp_setting}
\subsubsection{Datasets}

To validate the effectiveness of our proposed Refiner framework, we conduct experiments on two widely adopted public datasets and one real-world large-scale industrial dataset. The dataset statistics are summarized in Table~\ref{tab:dataset}.

\begin{table}[t]
\centering
\caption{Dataset statistics of the experimental datasets. ``User'', ``Item'', and ``Interaction'' denote the numbers of users, items, and user-item interactions, respectively. ``Avg. Length'' denotes the average length of user interaction sequences.}
\label{tab:dataset}
\setlength{\tabcolsep}{1mm}{
\resizebox{0.475\textwidth}{!}{
\begin{tabular}{c|cccc}
\toprule
\textbf{Datasets} & \textbf{\# User} & \textbf{\# Item} & \textbf{\# Interaction} & \textbf{Avg. Length} \\ \hline
\textbf{Steam} & 33,327 & 14,377  & 1.4 million & 43 \\
\textbf{MovieLens10M} & 3,558 & 10,680  & 10 million & 506 \\ 
\textbf{Industrial} & 70 million & 4 million  & 2 billion  & 1363 \\ \bottomrule
\end{tabular}
}}
\end{table}

\noindent $\bullet \quad${\textbf{Public Datasets}}. We choose two publicly available real-world datasets for experiments. (1) \textbf{Steam}~\cite{kang2018self} contains extensive user interactions with video games on the Steam platform. (2) \textbf{MovieLens10M}~\cite{harper2015movielens} is a widely used dataset for movie recommendation, containing diverse and complex user-movie interactions. For both datasets, we chronologically sort user interactions based on timestamps and partition them into training, validation, and test sets with a ratio of 8:1:1~\cite{lin2025order,wang2023causal,pi2019practice}. In all three splits, each positive interaction is paired with one randomly sampled negative item following previous work~\cite{wang2025scaling,pi2019practice}, yielding a candidate set of $K=2$; the same construction is used for training and evaluation. We use only item IDs and item category IDs as the discrete features for both user interaction sequences and candidate items.

\noindent $\bullet \quad${\textbf{Industry Dataset}}. To validate the effectiveness of our method in real-world recommendation scenarios, we construct an industrial-scale dataset from exposure logs of a deployed industrial recommender system following previous work~\cite{han2025mtgr}. The request-level candidate-set sizes follow the logged production pipeline rather than a Refiner-specific fixed $K$. For offline experiments, we use one full day of data for training and a portion of the subsequent day's data for testing. The same candidate construction is used for offline training and evaluation. For online experiments, we use a 14-day dataset, where the first 7 days are used to warm up Refiner and the subsequent 7 days serve as the online observation period while Refiner is updated following the MTGR update schedule. It is worth noting that user interaction history is further divided into long-term historical interactions and short-term real-time interactions in this industrial dataset. However, this distinction does not affect the application of our method.

\subsubsection{Backbone and Baseline} 
We implement our proposed Refiner framework on \textbf{MTGR}~\cite{han2025mtgr}, a deployed industrial GR model that has demonstrated significant business impact in production, to validate its practical utility and effectiveness. To further demonstrate the robustness of Refiner, we introduce \textbf{SASRec-ranking}, a ranking backbone adapted from SASRec~\cite{kang2018self}. Specifically, we implement the encoding of the aforementioned features and modify the original attention mask into a candidate mask to adapt SASRec to the ranking stage. We further include several representative traditional point-wise ranking models, \textbf{xDeepFM}~\cite{lian2018xdeepfm} and \textbf{DIN}~\cite{zhou2018deep}, to validate the effectiveness of MTGR and SASRec-ranking. We include \textbf{ReaRec}~\cite{tang2025think} as a representative latent reasoning baseline originally designed for retrieval. ReaRec models user representation learning as a latent reasoning process over item sequences, where iterative reasoning refines the user representation for item matching. Since it does not explicitly model multiple candidate items in one forward pass, we adapt it to ranking by appending candidate items to the input sequence and extending its attention mechanism for multi-candidate scoring without information leakage across candidates, enabling a fair comparison in the ranking scenario.

\subsubsection{Evaluation Metrics}
We focus on two prediction tasks, CTR and CTCVR, on the industrial dataset following~\cite{han2025mtgr}, and frame the ranking task as CTR prediction on the public datasets~\cite{wang2025scaling}. We adopt AUC and Group AUC (GAUC) to evaluate model performance. GAUC is calculated by averaging user-level AUC, offering a more detailed assessment of ranking quality.
\begin{table*}[!t]
\renewcommand{\arraystretch}{1.2}
\belowrulesep=0pt
\aboverulesep=0pt
\setlength{\abovecaptionskip}{0.05cm}
\setlength{\belowcaptionskip}{0.2cm}
\caption{CTR performance comparison on two public datasets. Refiner and ReaRec are instantiated on both SASRec-ranking and MTGR backbones. Performance improvements are indicated by ``$\uparrow$'' in (Improv.). $^{*}$ indicates that the improvement is statistically significant (\(p\)-value < 0.05) over the best baseline.}
{
\label{tab:main_table_public}
\setlength{\tabcolsep}{1.5mm}{
\resizebox{0.88\textwidth}{!}{
\begin{tabular}{c|cccc|cccc}
\toprule
  \multirow{2}{*}{\textbf{Method}} & \multicolumn{4}{c|}{\textbf{Steam}} & \multicolumn{4}{c}{\textbf{MovieLens10M}} \\ 
  \cmidrule(lr){2-5} \cmidrule(lr){6-9}
  & \multicolumn{2}{c}{\textbf{AUC}} & \multicolumn{2}{c|}{\textbf{GAUC}} & \multicolumn{2}{c}{\textbf{AUC}} & \multicolumn{2}{c}{\textbf{GAUC}} \\
  \cmidrule(lr){2-3} \cmidrule(lr){4-5} \cmidrule(lr){6-7} \cmidrule(lr){8-9}
  & \textbf{Mean} & \textbf{Std} & \textbf{Mean} & \textbf{Std} & \textbf{Mean} & \textbf{Std} & \textbf{Mean} & \textbf{Std} \\ \midrule
  \textbf{xDeepFM} & 0.6560 & 0.0013 & 0.6501 & 0.0015 & 0.6433 & 0.0015 & 0.5919 & 0.0011 \\ 
  \textbf{DIN} & 0.7522 & 0.0017 & 0.7261 & 0.0013 & 0.6583 & 0.0016 & 0.6369 & 0.0019 \\ 
  \midrule
  \textbf{SASRec-ranking} & 0.7450 & 0.0015 & 0.7342 & 0.0012 & 0.6935 & 0.0018 & 0.6651 & 0.0019 \\ 
  \textbf{SASRec-ranking+ReaRec} & 0.7494 & 0.0011 & 0.7394 & 0.0017 & 0.6988 & 0.0014 & 0.6705 & 0.0013 \\ 
  \rowcolor{ours_blue} \textbf{SASRec-ranking+Refiner} & \textbf{0.7511$^{*}$} & 0.0016 & \textbf{0.7415$^{*}$} & 0.0014 & \textbf{0.7022$^{*}$} & 0.0010 & \textbf{0.6742$^{*}$} & 0.0015 \\ 
  \midrule
  \textbf{Improv.} & ($\uparrow 0.8\%$) & -- & ($\uparrow 1.0\%$) & -- & ($\uparrow 1.3\%$) & -- & ($\uparrow 1.4\%$) & -- \\ \midrule
  \textbf{MTGR} & 0.7698 & 0.0025 & 0.7397 & 0.0023 & 0.7674 & 0.0021 & 0.6755 & 0.0025 \\ 
  \textbf{MTGR+ReaRec} & 0.7785 & 0.0028 & 0.7443 & 0.0021 & 0.7719 & 0.0024 & 0.6827 & 0.0027 \\ 
  \rowcolor{ours_blue} \textbf{MTGR+Refiner} & \textbf{0.7872$^{*}$} & 0.0027 & \textbf{0.7594$^{*}$} & 0.0029 & \textbf{0.7864$^{*}$} & 0.0022 & \textbf{0.7043$^{*}$} & 0.0029 \\ 
  \midrule
  \textbf{Improv.} & ($\uparrow 2.3\%$) & -- & ($\uparrow 2.7\%$) & -- & ($\uparrow 2.5\%$) & -- & ($\uparrow 4.3\%$) & -- \\ \bottomrule
\end{tabular}
}}}
\end{table*}

\begin{table*}[!t]
\renewcommand{\arraystretch}{1.2}
\belowrulesep=0pt
\aboverulesep=0pt
\setlength{\abovecaptionskip}{0.05cm}
\setlength{\belowcaptionskip}{0.2cm}
\centering
\caption{CTR and CTCVR performance comparison on the industrial dataset. Refiner is compared with the latent reasoning baseline ReaRec. Performance improvements are indicated by ``$\uparrow$'' in (Improv.). $^{*}$ indicates that the improvement is statistically significant (\(p\)-value < 0.05) over the best baseline.}
\label{tab:main_table_industry}
\setlength{\tabcolsep}{4pt}
\resizebox{0.78\textwidth}{!}{
\begin{tabular}{c|cccc|cccc}
\toprule
  \multirow{2}{*}{\textbf{Method}} & \multicolumn{4}{c|}{\textbf{CTR}} & \multicolumn{4}{c}{\textbf{CTCVR}} \\ 
  \cmidrule(lr){2-5} \cmidrule(lr){6-9}
  & \multicolumn{2}{c}{\textbf{AUC}} & \multicolumn{2}{c|}{\textbf{GAUC}} & \multicolumn{2}{c}{\textbf{AUC}} & \multicolumn{2}{c}{\textbf{GAUC}} \\
  \cmidrule(lr){2-3} \cmidrule(lr){4-5} \cmidrule(lr){6-7} \cmidrule(lr){8-9}
  & \textbf{Mean} & \textbf{Std} & \textbf{Mean} & \textbf{Std} & \textbf{Mean} & \textbf{Std} & \textbf{Mean} & \textbf{Std} \\ \midrule
  
  \textbf{MTGR} & 0.7404 & 0.0003 & 0.6909 & 0.0002 & 0.8553 & 0.0002 & 0.7046 & 0.0003 \\ 
  \textbf{MTGR+ReaRec} & 0.7410 & 0.0003 & 0.6913 & 0.0002 & 0.8567 & 0.0003 & 0.7054 & 0.0003 \\ 
  \rowcolor{ours_blue} \textbf{MTGR+Refiner} & \textbf{0.7417$^{*}$} & 0.0004 & \textbf{0.6920$^{*}$} & 0.0003 & \textbf{0.8574$^{*}$} & 0.0003 & \textbf{0.7061$^{*}$} & 0.0003 \\ 
  \midrule
  \textbf{Improv.} & ($\uparrow 1.8\text{\textperthousand}$) & -- & ($\uparrow 1.6\text{\textperthousand}$) & -- & ($\uparrow 2.5\text{\textperthousand}$) & -- & ($\uparrow 2.1\text{\textperthousand}$) & -- \\ \bottomrule
\end{tabular}}
\end{table*}
\subsubsection{Implementation Details}
All experiments use NVIDIA A100 GPUs and 3 random seeds. For MTGR in the offline industrial experiments, we adopt the architecture with 3 Transformer blocks, 2 attention heads, and an embedding dimension of 512 following~\cite{han2025mtgr}. We use smaller model sizes for MTGR and SASRec-ranking on the public datasets to avoid overfitting. For traditional CTR models, we adjust the embedding dimension among $\{32, 64, 128, 256, 512\}$. The learning rate $\eta$ for MTGR and traditional recommender models is tuned over $\{1\times10^{-4}, 5\times10^{-4}, 1\times10^{-3}\}$. For our proposed Refiner framework, we perform a hyperparameter search for the LoRA rank $r$ over $\{32, 64, 128\}$ and its corresponding scaling factor $\alpha$ over $\{64.0, 128.0, 256.0\}$. The number of history queries $N_{\text{HQ}}$ is selected from $\{0, 1, 2, 4, 8, 16\}$. Regarding the online experiments, we additionally construct an incremental update framework. This framework leverages a warm-start strategy, where each day's Refiner model is initialized from the previous day's Refiner checkpoint and the current day's MTGR checkpoint. During each daily Refiner update, the corresponding MTGR checkpoint is kept frozen, while only the additional Refiner parameters are optimized on the current day's data. In this way, Refiner is updated following the MTGR update schedule. A detailed explanation of the online update framework is provided in Section~4.3.5.

\subsection{Overall Performance (RQ1)}\label{sec:overall_performance}
We conduct extensive comparisons among traditional ranking models, GR backbones, and latent reasoning methods instantiated on both MTGR and SASRec-ranking. The experimental results on public datasets are presented in Table~\ref{tab:main_table_public}, while the industrial dataset results are shown in Table~\ref{tab:main_table_industry}. In the main experiments, the total number of reasoning steps is fixed to 1, and the impact of varying this hyperparameter is further analyzed in Section~\ref{sec:RS}. From Table~\ref{tab:main_table_public} and Table~\ref{tab:main_table_industry}, we have the following key observations:

\begin{itemize}[leftmargin=*]
    \item On public datasets, MTGR and SASRec-ranking generally achieve substantial performance gains over traditional recommender systems. Specifically, as shown in Table~\ref{tab:main_table_public}, MTGR substantially outperforms xDeepFM and DIN on both Steam and MovieLens10M, demonstrating the effectiveness of GR backbones for ranking tasks.
    
    \item Both evaluated latent-reasoning extensions, \ie ReaRec and Refiner, improve their corresponding direct-prediction backbones across the public and industrial datasets. In industrial practice, an improvement of 1\textperthousand  is generally considered meaningful. For example, on the CTR GAUC metric, the observed gain yields a paired t-test \(p\)-value of 0.008, confirming its statistical significance. These results show that the evaluated extensions provide gains beyond direct prediction in our settings.
    
    \item Refiner consistently achieves higher AUC and GAUC than the adapted ReaRec baseline on both public and industrial datasets. Together with the component analysis in Section~\ref{sec:abl}, these results support the effectiveness of Refiner's ranking-oriented design in the evaluated settings.
\end{itemize}
\noindent\textbf{Remark.} The smaller numerical gains on the industrial dataset are likely due to the fact that Refiner is built on top of an already deployed recommender in a large-scale production environment, where the remaining room for improvement is naturally more limited than on public benchmarks. Nevertheless, these consistent offline gains remain meaningful in practice, and their practical value is further supported by the online results in Section~\ref{sec:online_experiment}.

\subsection{Further Analysis}
In this section, we conduct a more detailed investigation of the proposed Refiner framework. Specifically, in Section~\ref{sec:abl}, we perform ablation studies to examine the impact of removing individual components (\textbf{RQ2}). In Section~\ref{sec:ga}, we analyze Refiner's performance across different user behavior sequence lengths and item popularity groups in the test set to demonstrate its robustness (\textbf{RQ3}). Furthermore, Sections~\ref{sec:NHQ} and~\ref{sec:RS} investigate the effects of the number of history queries $N_{\text{HQ}}$ and reasoning steps $R_{\text{steps}}$ on both model performance and inference-time overhead (\textbf{RQ4}), providing guidance for applying Refiner in real-world systems.
\begin{table}[t]
\renewcommand{\arraystretch}{1.2}
\caption{Ablation study of Refiner on the Steam dataset. Performance declines are indicated by ``$\downarrow$'' in (Dec.).}
{
\label{tab:ablation}
\renewcommand{\arraystretch}{1.2}
\setlength{\tabcolsep}{1.5mm}{
\resizebox{0.45\textwidth}{!}{
{
\begin{tabular}{c|cccc}
\toprule
  \multirow{2}{*}{\textbf{Method}} & \multicolumn{4}{c}{\textbf{CTR}} \\ 
  \cline{2-5}
   & \textbf{AUC} & \textbf{(Dec.)}  & \textbf{GAUC} & \textbf{(Dec.)}  \\ 
  \hline
  \textbf{Refiner} & \textbf{0.7872} & - & \textbf{0.7594} & -   \\ 
  
  \textbf{w/o H.Q.} 
  & 0.7764 
  & ($\downarrow1.4\%$) 
  & 0.7495 
  & ($\downarrow1.3\%$) \\ 
  
  \textbf{w/o LoRA} 
  & 0.7841 
  & ($\downarrow0.4\%$) 
  & 0.7413 
  & ($\downarrow2.4\%$) \\
  
  \textbf{w/o Residual} 
  & 0.7763 
  & ($\downarrow1.4\%$) 
  & 0.7523 
  & ($\downarrow0.9\%$) \\
  
  \textbf{End-to-End} 
  & 0.7650 
  & ($\downarrow2.8\%$) 
  & 0.7316 
  & ($\downarrow3.7\%$) \\ 
  
  \bottomrule
\end{tabular}
}}}}
\end{table}
\subsubsection{\textbf{Ablation Study (RQ2)}}\label{sec:abl}

We systematically evaluate the effectiveness of individual components in the Refiner framework on the Steam dataset through ablation studies. Specifically, we examine the following variants: removing history queries (w/o H.Q.), removing the LoRA adaptation in $\mathcal{T}_{\text{reasoning}}$ (w/o LoRA), removing residual reasoning (w/o Residual), \ie replacing the residual refinement objective with direct prediction supervision using BCE loss at each reasoning step, and an End-to-End training framework that directly optimizes the initialized model with both $\mathcal{L}_{\text{Base}}$ and $\mathcal{L}_{\text{reasoning}}$ without the additional LoRA module and $\mathcal{P}_{\text{reasoning}}$. The corresponding results are summarized in Table~\ref{tab:ablation}.

\begin{figure*}[!t]
\centering
\includegraphics[scale=0.64
]{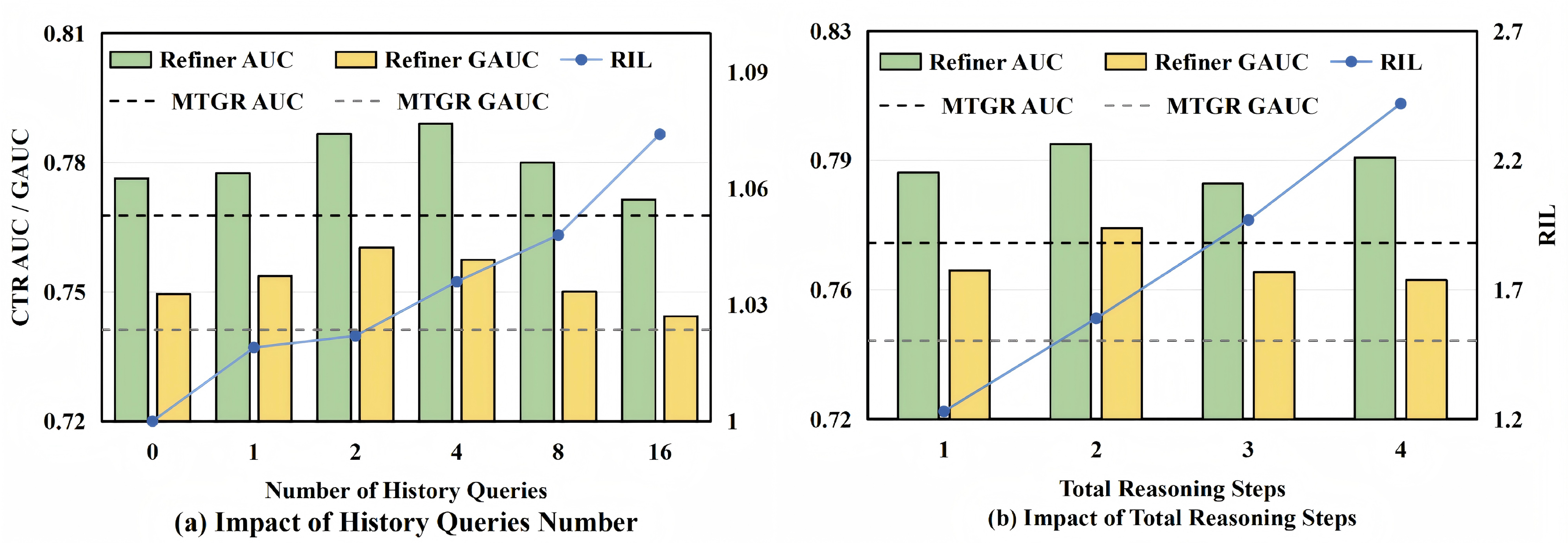}
\caption{The performance (CTR AUC/GAUC) and relative inference latency (RIL) variation trend of Refiner under different (a) number of history queries and (b) total reasoning steps on Steam dataset.}
\label{fig:NR}
\end{figure*}

The experimental results in Table~\ref{tab:ablation} show that: \textbf{(1)} removing any evaluated component leads to performance degradation, supporting the effectiveness of the complete Refiner framework. \textbf{(2)} Removing history queries degrades performance, suggesting that compact global user context contributes to candidate-specific reasoning in the evaluated setting. \textbf{(3)} Under the evaluated reasoning architecture, replacing residual refinement with direct BCE supervision at each step yields lower performance, supporting the cumulative residual formulation over this direct-prediction alternative. \textbf{(4)} The End-to-End variant performs worst among the evaluated variants, suggesting that directly optimizing the full model may weaken the functional separation between base prediction and reasoning refinement.

\begin{figure}[t]
\setlength{\abovecaptionskip}{0.2 cm}
\setlength{\belowcaptionskip}{-0.2cm}

\centering
\includegraphics[scale=0.37]{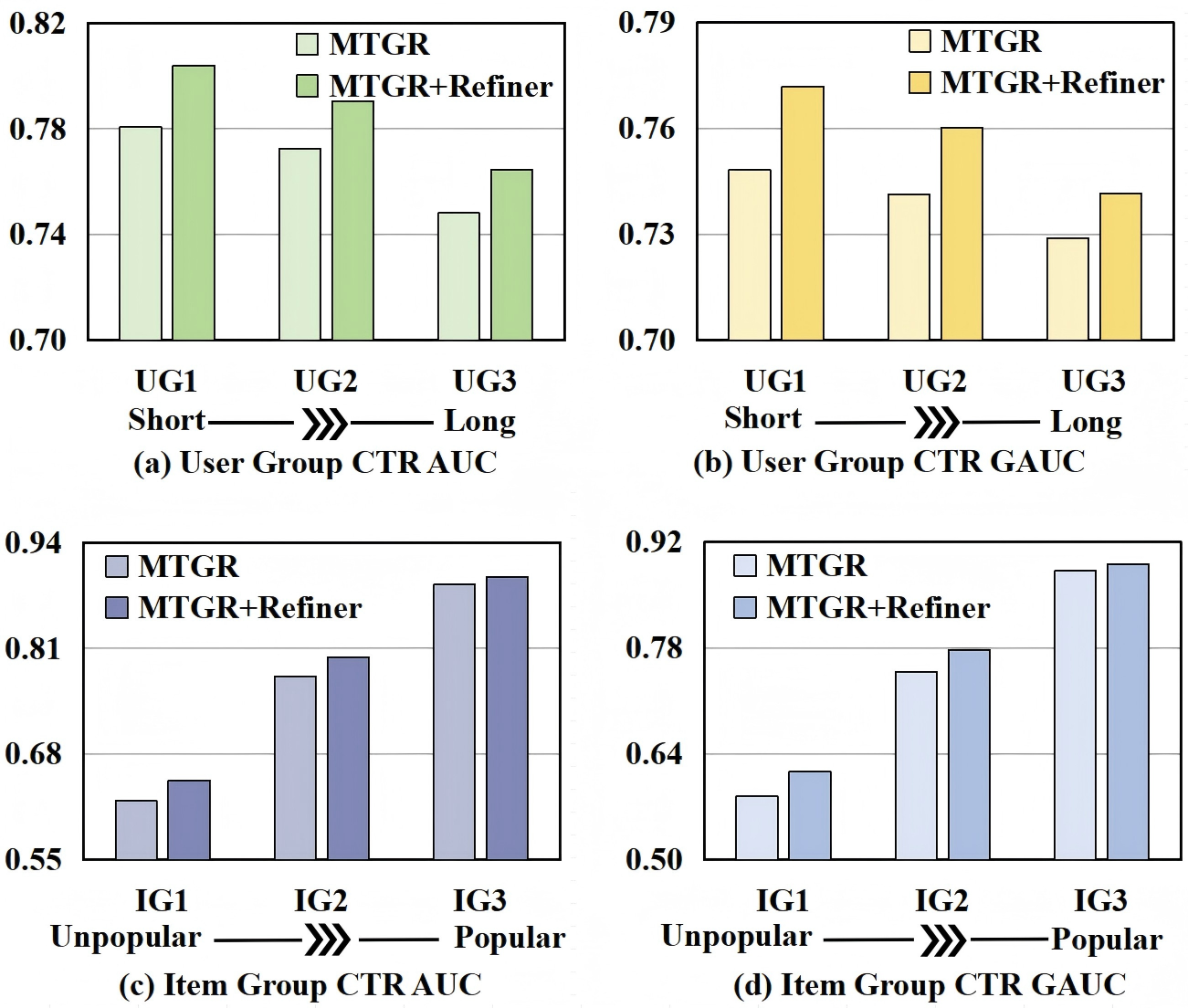}
\caption{Robustness study w.r.t.\ different user and item subgroups on Steam dataset. UG and IG denote User and Item Group, respectively, where higher group numbers indicate longer sequences and more popular items.}
\label{fig:group}
\end{figure}
\subsubsection{\textbf{Group Analysis (RQ3)}}\label{sec:ga}
To evaluate the robustness of the proposed Refiner framework, we conduct a group-wise analysis by partitioning users and items into distinct subgroups. Specifically, users are categorized into three equal-sized groups based on their behavior sequence length, with higher group indices corresponding to longer sequences. Similarly, items are grouped into three categories based on interaction frequency, where higher indices denote more popular items. Each item group is balanced to contain an equal number of samples following established practice~\cite{tang2025think}. The results are depicted in Figure~\ref{fig:group}.

From Figure~\ref{fig:group}, we make the following observations: \textbf{(1)} As user sequence length increases, MTGR's performance gradually declines. This phenomenon may stem from the fact that longer user histories often involve more complex behavioral patterns and more long-tail items, which increase the difficulty of ranking. Refiner consistently improves performance across all user groups, demonstrating its robustness in handling users with different history lengths through candidate-specific latent refinement conditioned on user context. \textbf{(2)} MTGR's performance deteriorates as item popularity decreases. This trend is consistent with the general difficulty of modeling cold-start or less frequent items in recommender systems. Refiner consistently improves performance across all item groups, with the most pronounced gains on the least popular item group. This suggests that Refiner is particularly helpful when the base recommendation model has less reliable item representations and there is more room for refinement. For high-popularity items, the backbone model can already capture strong collaborative signals from sufficient interaction data, leaving less room for additional refinement. As a result, the relative gains from Refiner are smaller on the most popular items.

\subsubsection{\textbf{Impact of History Queries Number (RQ4)}}\label{sec:NHQ}
To systematically investigate the impact of the number of history queries $N_{\text{HQ}}$ on both model performance and inference-time overhead, we conduct a sensitivity study by varying $N_{\text{HQ}}\in\{0,1,2,4,8,16\}$ while fixing the number of reasoning steps to 1 and keeping all other parameters unchanged. The corresponding performance metrics (AUC and GAUC) and relative inference latency (RIL) are presented in Figure~\ref{fig:NR}(a). Here, RIL is computed relative to the baseline configuration with $N_{\text{HQ}}=0$ and one reasoning step, providing a normalized measure of the additional inference overhead introduced by different history-query settings.

From Figure~\ref{fig:NR}(a), we observe two key findings: \textbf{(1)} A moderate number of history queries (e.g., $N_{\text{HQ}}=2$ or $4$) achieves the best performance. When $N_{\text{HQ}}$ is too small, the reasoning process has limited access to condensed global user context, which weakens refinement quality. In contrast, excessively large $N_{\text{HQ}}$ may introduce redundant history signals and make the refinement process less focused. \textbf{(2)} Introducing a small number of history queries incurs limited relative overhead in the current implementation, with RIL remaining below 1.1 compared to the baseline. These results support the effectiveness of history queries in the evaluated setting, showing that a small number of compact user-context tokens provides useful global information with limited relative overhead.
\subsubsection{\textbf{Impact of Total Reasoning Steps (RQ4)}}\label{sec:RS}
In this section, we investigate the impact of the total number of reasoning steps $R_{\text{steps}}$ on both model performance and inference-time overhead. Specifically, we vary $R_{\text{steps}}\in\{1,2,3,4\}$ during both training and inference, and present the results in Figure~\ref{fig:NR}(b). Here, RIL is calculated relative to the one-shot prediction of the MTGR backbone.
\begin{figure}[t]

\centering
\centering
\includegraphics[scale=0.63
]{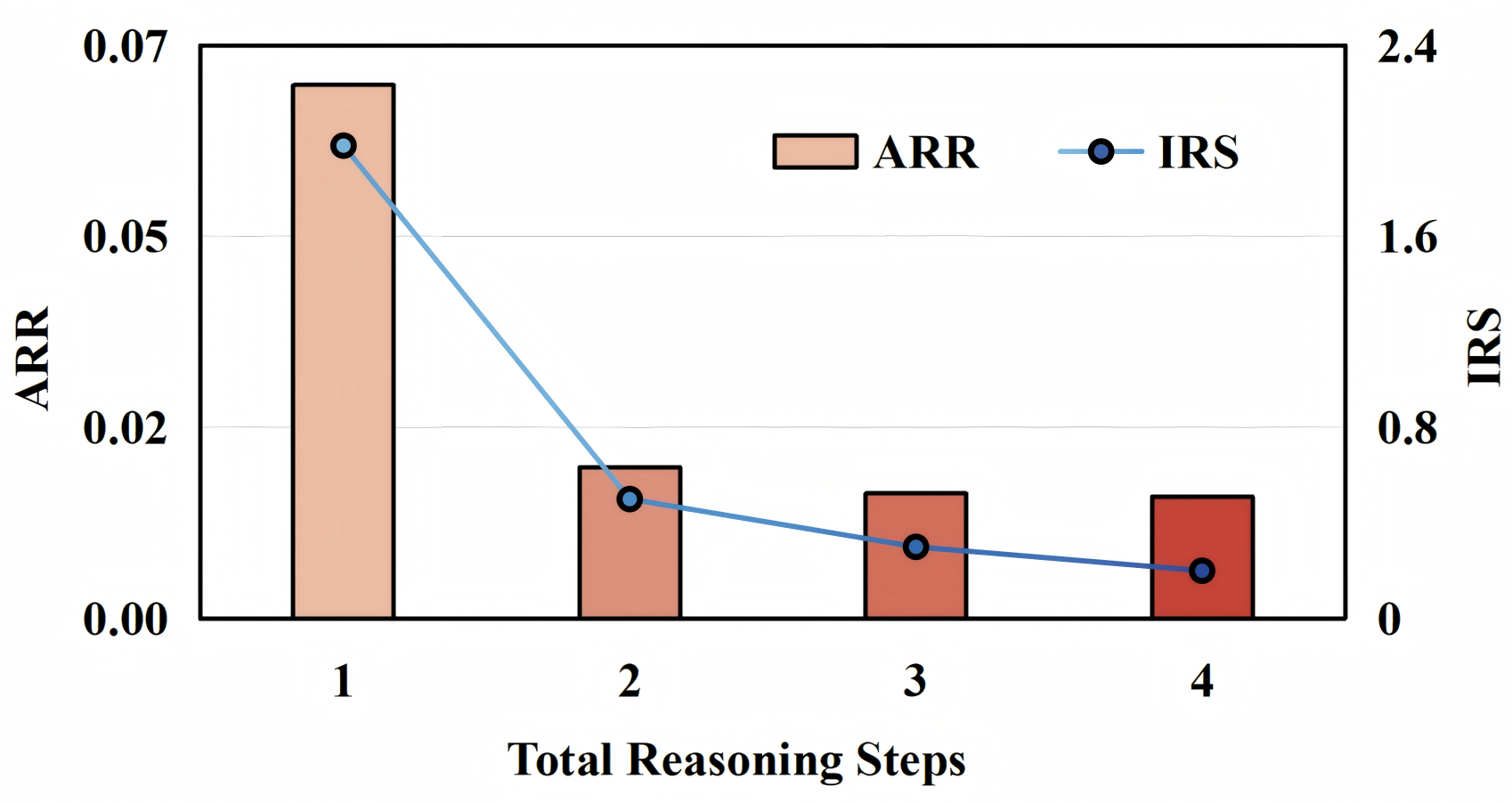}
\caption{The average relative ratio (ARR) of the absolute residual-score magnitude at each reasoning step to the absolute direct-prediction score magnitude, together with the incremental performance improvement per step relative to the preceding step (IRS), on the Steam dataset.}
\label{fig:rel_abs}
\end{figure}
From Figure~\ref{fig:NR}(b), we observe that Refiner achieves clear performance improvements over the base model when using a small number of reasoning steps. This result indicates that a limited amount of multi-step latent refinement is already effective for improving ranking quality beyond direct one-shot prediction. However, as the number of reasoning steps continues to increase, the additional gains quickly diminish, a trend that is also observed in~\cite{tang2025think,tang2026parallel}. To further analyze this behavior, we visualize the average relative ratio (ARR) between the absolute residual score magnitude at each reasoning step and the absolute magnitude of the initial direct prediction scores, together with the incremental performance improvement per step relative to the previous step (IRS) in the 4-step reasoning model, as shown in Figure~\ref{fig:rel_abs}. The results show that the magnitude of residual scores progressively decreases as the reasoning process deepens, indicating that later steps contribute only limited additional adjustments to the initial predictions. Combined with the steadily increasing inference latency, these findings suggest that 1 or 2 reasoning steps are sufficient in our evaluated settings to achieve a favorable trade-off between performance gains and inference-time overhead.

\noindent\textbf{Remark.} The reported RIL reflects the current implementation of Refiner. Since the additional reasoning passes have not undergone dedicated serving-level optimization, these relative measurements should not be interpreted as a general end-to-end efficiency comparison with fully optimized production rankers.

\begin{figure}[t]
\centering
\centering
\includegraphics[scale=0.235
]{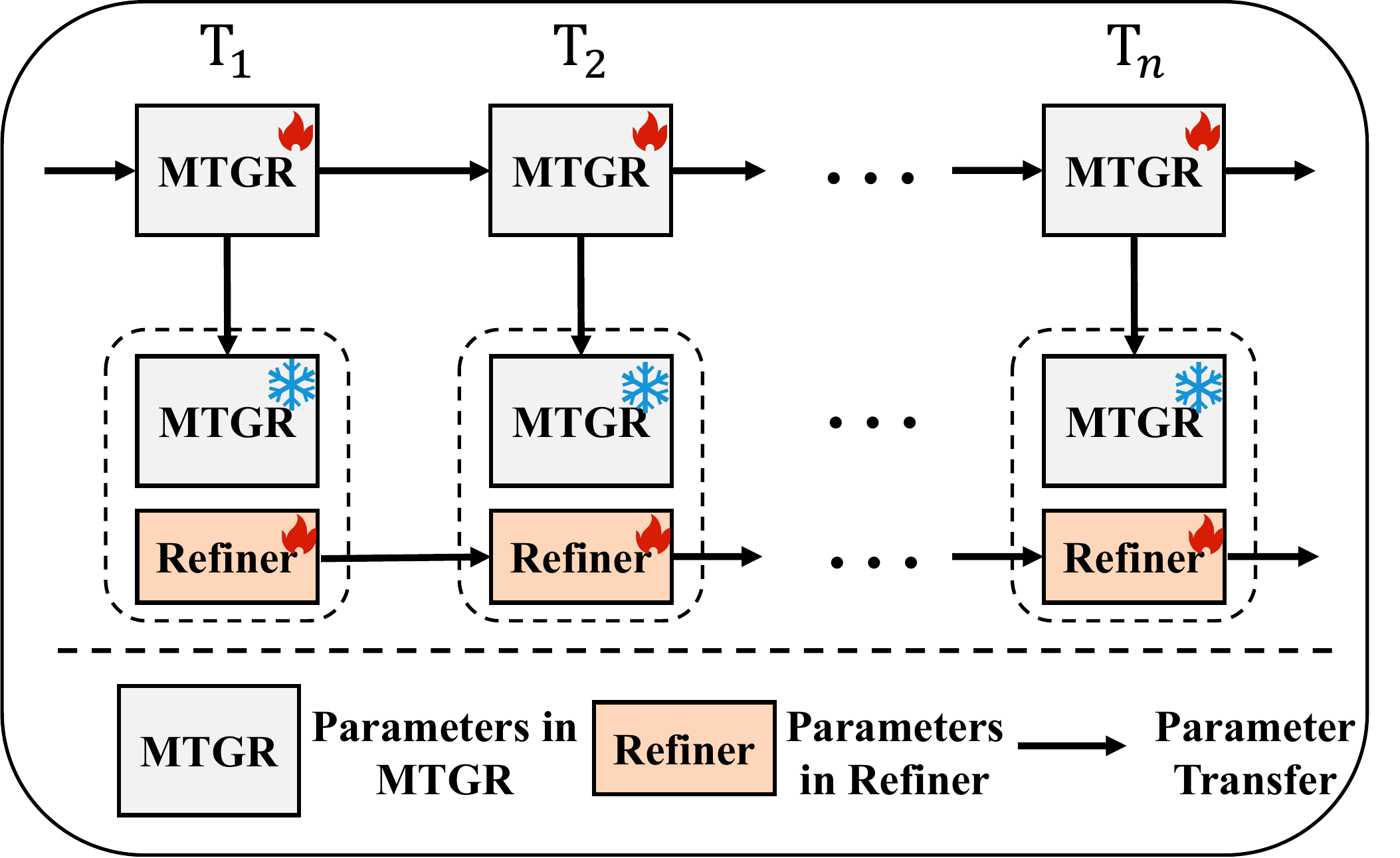}
\caption{Incremental update framework for online deployment. At each iteration, Refiner loads and freezes the latest MTGR checkpoint, initializes its trainable parameters from the previous Refiner checkpoint, and updates only these parameters on the latest data. This keeps Refiner aligned with MTGR updates without full-model retraining.}
\label{fig:incre}
\end{figure}
\subsubsection{\textbf{Online Experiments (RQ5)}}\label{sec:online_experiment}
To further evaluate the practical utility of Refiner in a real-world industrial setting, we deploy Refiner on top of the 15-layer online MTGR model in a large-scale online recommendation platform. Online candidates are supplied by the upstream retrieval stage, and the same pipeline-defined candidate construction is used for training and evaluation. In the online setting, Refiner uses a single reasoning step due to latency considerations, following the offline trade-off between effectiveness and inference-time overhead. We conduct an online A/B test on a 1\% production traffic bucket, with the production MTGR model as the control and MTGR+Refiner as the treatment. The experimental traffic covers millions of exposures per day, providing a reliable basis for online evaluation.
To align Refiner with the daily update pipeline of the deployed system, we build an incremental update framework, as illustrated in Figure~\ref{fig:incre}. Specifically, each day's Refiner is initialized from the previous day's Refiner checkpoint together with the current day's MTGR checkpoint. During each daily Refiner update, the corresponding MTGR checkpoint is kept frozen, while only the additional parameters of Refiner are optimized on the current day's data. Following this incremental framework, we first warm up Refiner for 7 days and then conduct A/B testing over a subsequent 7-day observation window, during which Refiner remains updated following the online MTGR model schedule.

As shown in Table~\ref{tab:ol}, Refiner achieves consistent online gains over MTGR during the 7-day experimental window, with PV-CTR increasing by 0.75\%, UV-CTCVR increasing by 0.21\%, and GMV increasing by 0.46\%. These results demonstrate that Refiner can be effectively integrated into a real-world recommender system and deliver meaningful business improvements on top of an already deployed strong ranker. For completeness, we provide the daily performance trend over the 7-day observation window in Appendix~\ref{app:online_daily_trend}, which further illustrates the stability of the online gains.
\begin{table}[H]
    \vspace{-0.7em}
    \centering
    \caption{Seven-day online A/B-test improvements of Refiner over MTGR. PV, UV, and GMV denote page views, unique visitors, and gross merchandise volume, respectively.}
    \vspace{-1em}
    \label{tab:ol}
    \resizebox{\linewidth}{!}{
    \begin{tabular}{*{6}{c}}
        \toprule
        \textbf{Metric} & PV-CTR & UV-CTR & UV-CTCVR & Order-Number & GMV  \\
        \midrule
        \textbf{Impr.} & +0.75\% & +0.12\% & +0.21\% & +0.19\%  & +0.46\%   \\
        \bottomrule
    \end{tabular}
    }
    \vspace{-0.5em}
\end{table}

\section{Related Work}

\subsection{Generative Recommendation}
Generative Recommendation (GR)~\cite{zhai2024actions,han2025mtgr,rajput2023recommender,huang2025towards,deng2025onerec,wang2024learnable,song2026mtfm} has emerged as a promising paradigm by reformulating recommendation as a sequence transduction problem. Existing GR research can be broadly divided into two categories based on tokenization strategies and training objectives. The first category~\cite{rajput2023recommender,deng2025onerec,wang2024learnable} emphasizes semantic item identification, where items are tokenized into discrete semantic codes for autoregressive generalization. Representative works such as TIGER~\cite{rajput2023recommender} leverage RQ-VAE~\cite{lee2022autoregressive} to tokenize items into hierarchical semantic IDs for autoregressive item prediction. The second category~\cite{zhai2024actions,han2025mtgr,huang2025towards} focuses on encoding item features into dense embeddings while adopting discriminative training objectives. Representative works in this line have demonstrated strong performance in large-scale recommendation settings. In ranking variants, user context and explicit candidate items are jointly encoded to produce candidate scores through parallel processing. Our work builds on this paradigm and introduces latent reasoning for progressive score refinement.

\subsection{Reasoning for Recommendation}
Reasoning-enhanced recommendation broadly follows explicit or latent paradigms. Explicit methods~\cite{fang2025reason4rec,bismay2024reasoningrec,gu2025r,yu2025thinkrec} use LLMs~\cite{touvron2023llama,bai2023qwen} to generate textual reasoning chains, but incur substantial inference overhead. Latent methods instead perform multi-step computation in continuous hidden spaces~\cite{tang2025think,dai2026onepiece,liu2025lares,zhang2025reinforced,tang2026parallel,lin2026care,yang2026beyond,shan2026calr}; for example, LatentR$^3$~\cite{zhang2025reinforced} optimizes compact latent reasoning in LLM-based recommenders through reinforcement learning.

Within the latent paradigm, existing methods target different recommendation interfaces and learning objectives. CARE~\cite{lin2026care} refines autoregressive semantic-ID generation, whereas ReaRec, CaLR, and related methods~\cite{tang2025think,shan2026calr,liu2025lares,tang2026parallel,yang2026beyond} refine user representations for next-item recommendation. OnePiece~\cite{dai2026onepiece} covers retrieval and ranking through grouped candidate interaction and progressive supervision. Its full ranking configuration trains a set-wise backbone end to end, allows interaction among candidates, and uses contextual inputs defined by the business setting. Refiner instead formulates each reasoning step as an additive correction to candidate scores, while preserving isolated candidate processing and training only lightweight modules.

\section{Conclusion and Future Work}

This work investigates latent reasoning for fine-grained candidate scoring in generative ranking. To address the key design challenges, \textbf{Refiner} incorporates three components: \textbf{(1)} a residual reasoning mechanism that accumulates score corrections across reasoning steps to align latent computation with the ranking objective; \textbf{(2)} history queries that retain compact global user context during candidate-specific reasoning; and \textbf{(3)} a parameter-efficient fine-tuning framework that adapts established ranking backbones without retraining the full model. Experiments on two public datasets, a large-scale offline industrial dataset, and an online A/B test show consistent improvements over the evaluated baselines, while further analyses indicate gains across different user and item groups.

For future work, one promising direction is adaptive reasoning depth selection. Refiner currently applies a fixed reasoning depth to all instances, although simple cases may require fewer refinement steps while harder cases may benefit from additional computation. Future work may investigate adaptive depth allocation or stopping mechanisms that balance refinement quality and inference cost.

\textbf{Limitations.} The evaluation has two main limitations. \textbf{(1)} Differences in task interfaces and backbone requirements restrict the number of directly comparable methods for latent reasoning. \textbf{(2)} A fair comparison with OnePiece~\cite{dai2026onepiece} requires a backbone that supports interaction among candidates and contextual inputs defined by its business setting, which are unavailable in our current production setup. Nevertheless, the industrial offline results and online A/B test provide evidence of Refiner's practical effectiveness in the evaluated production setting.

\begin{acks}
This work was supported by the \grantsponsor{GS202402KFLH01}{Open Innovation Program for jointly trained master's-to-doctoral students at the Institute of Computing Technology, Chinese Academy of Sciences}{} (\grantnum{GS202402KFLH01}{2024-02-KF-LH-01}).
\end{acks}

\appendix

\section{Additional Online Performance Analysis}

\subsection{Daily Online Performance Trend}
\label{app:online_daily_trend}
\begin{figure}[h]
    \centering
    \includegraphics[width=0.90\linewidth]{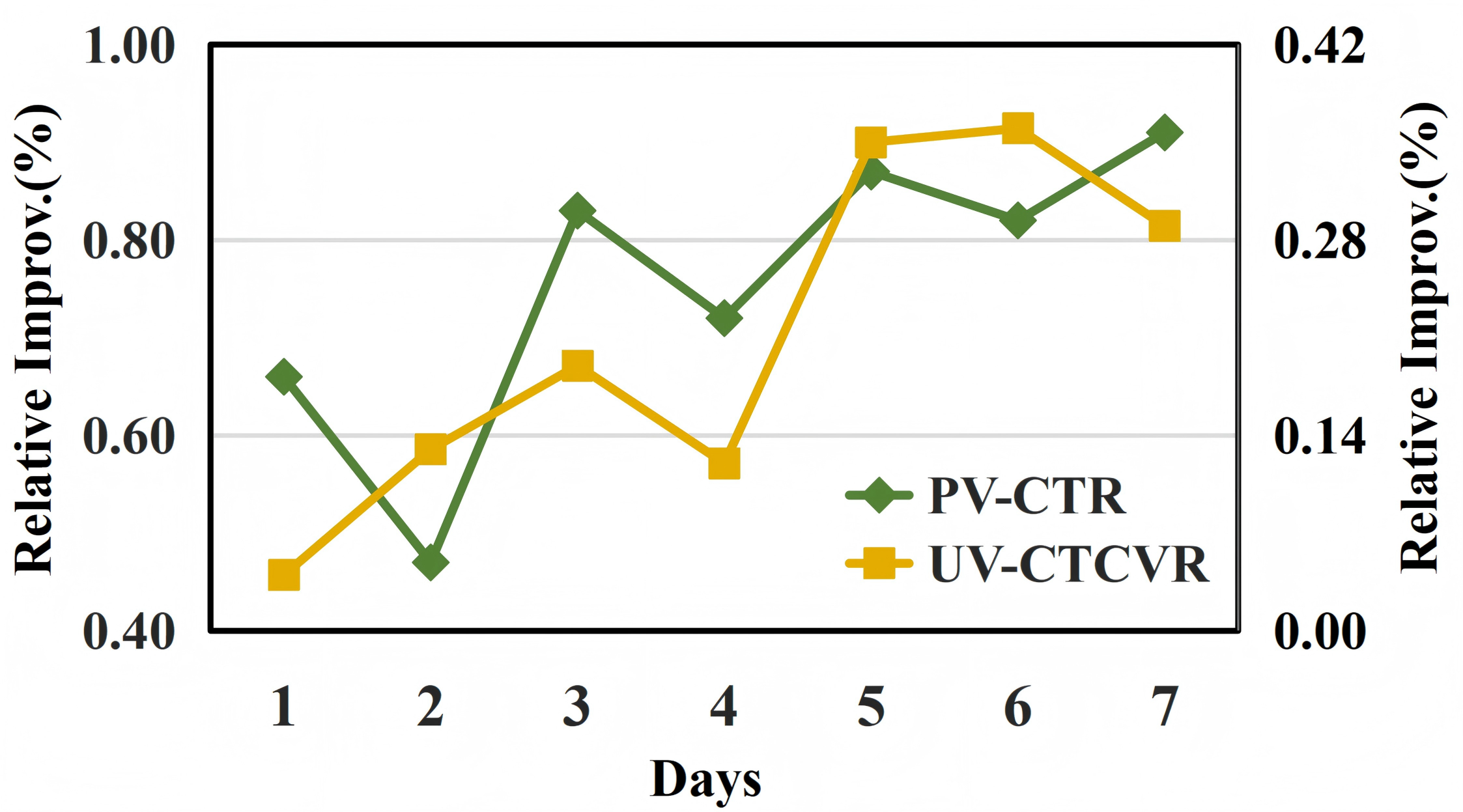}
    \caption{Daily relative improvements (\%) in PV-CTR (left axis) and UV-CTCVR (right axis) over the 7-day online observation window.}
    \label{fig:online_daily_trend}
\end{figure}

As shown in Figure~\ref{fig:online_daily_trend}, the relative gains in PV-CTR and UV-CTCVR remain positive throughout the 7-day observation window rather than being driven by a single day. Despite mild fluctuations under evolving traffic and model checkpoints, the overall trend supports the stability of Refiner under incremental updates.

\clearpage
\section*{GenAI Usage Disclosure}
\AddToHookNext{shipout/after}{\balance}
ChatGPT and OpenAI Codex assisted with language refinement, LaTeX checks, and code development and debugging. The authors reviewed and validated all outputs, conducted all experiments, and retain full responsibility.

\bibliographystyle{ACM-Reference-Format}
\bibliography{bibfile}

@inproceedings{kang2018self,
  author={Kang, Wang-Cheng and McAuley, Julian},
  title={Self-Attentive Sequential Recommendation},
  booktitle={2018 IEEE International Conference on Data Mining (ICDM)},
  year={2018},
  pages={197--206},
  doi={10.1109/ICDM.2018.00035}
}

@inproceedings{hou2022towards,
  author={Hou, Yupeng and Mu, Shanlei and Zhao, Wayne Xin and Li, Yaliang and Ding, Bolin and Wen, Ji-Rong},
  title={Towards Universal Sequence Representation Learning for Recommender Systems},
  booktitle={Proceedings of the 28th ACM SIGKDD Conference on Knowledge Discovery and Data Mining},
  year={2022},
  pages={585--593},
  numpages={9},
  series={KDD '22},
  publisher={Association for Computing Machinery},
  address={New York, NY, USA},
  location={Washington DC, USA},
  doi={10.1145/3534678.3539381}
}

@inproceedings{wang2025scaling,
  author={Wang, Chunqi and Wu, Bingchao and Chen, Zheng and Shen, Lei and Wang, Bing and Zeng, Xiaoyi},
  title={Scaling Transformers for Discriminative Recommendation via Generative Pretraining},
  booktitle={Proceedings of the 31st ACM SIGKDD Conference on Knowledge Discovery and Data Mining V.2},
  year={2025},
  pages={2893--2903},
  numpages={11},
  series={KDD '25},
  publisher={Association for Computing Machinery},
  address={New York, NY, USA},
  location={Toronto ON, Canada},
  doi={10.1145/3711896.3737117}
}

@inproceedings{han2025mtgr,
  author={Han, Ruidong and Yin, Bin and Chen, Shangyu and Jiang, He and Jiang, Fei and Li, Xiang and Ma, Chi and Huang, Mincong and Li, Xiaoguang and Jing, Chunzhen and Han, Yueming and Zhou, MengLei and Yu, Lei and Liu, Chuan and Lin, Wei},
  title={{MTGR}: Industrial-Scale Generative Recommendation Framework in Meituan},
  booktitle={Proceedings of the 34th ACM International Conference on Information and Knowledge Management},
  year={2025},
  pages={5731--5738},
  numpages={8},
  series={CIKM '25},
  publisher={Association for Computing Machinery},
  address={New York, NY, USA},
  location={Seoul, Republic of Korea},
  doi={10.1145/3746252.3761565}
}

@inproceedings{zhai2024actions,
  author={Zhai, Jiaqi and Liao, Lucy and Liu, Xing and Wang, Yueming and Li, Rui and Cao, Xuan and Gao, Leon and Gong, Zhaojie and Gu, Fangda and He, Jiayuan and Lu, Yinghai and Shi, Yu},
  title={Actions Speak Louder than Words: Trillion-Parameter Sequential Transducers for Generative Recommendations},
  booktitle={Proceedings of the 41st International Conference on Machine Learning},
  year={2024},
  pages={58484--58509},
  series={Proceedings of Machine Learning Research},
  volume={235},
  publisher={PMLR}
}

@inproceedings{huang2025towards,
  author={Huang, Yanhua and Chen, Yuqi and Cao, Xiong and Yang, Rui and Qi, Mingliang and Zhu, Yinghao and Han, Qingchang and Liu, Yaowei and Liu, Zhaoyu and Yao, Xuefeng and Jia, Yuting and Ma, Leilei and Zhang, Yinqi and Zhu, Taoyu and Zhang, Liujie and Chen, Lei and Chen, Weihang and Zhu, Min and Xu, Ruiwen and Zhang, Lei},
  title={Towards Large-scale Generative Ranking},
  booktitle={Proceedings of Second Workshop on Generative AI for Recommender Systems and Personalization},
  year={2025},
  numpages={6},
  series={KDD '25},
  publisher={Association for Computing Machinery},
  address={New York, NY, USA},
  location={Toronto, Canada}
}

@misc{deng2025onerec,
  author={Deng, Jiaxin and Wang, Shiyao and Cai, Kuo and Ren, Lejian and Hu, Qigen and Ding, Weifeng and Luo, Qiang and Zhou, Guorui},
  title={{OneRec}: Unifying Retrieve and Rank with Generative Recommender and Iterative Preference Alignment},
  year={2025},
  eprint={2502.18965},
  archivePrefix={arXiv},
  primaryClass={cs.IR}
}

@inproceedings{rajput2023recommender,
  author={Rajput, Shashank and Mehta, Nikhil and Singh, Anima and Hulikal Keshavan, Raghunandan and Vu, Trung and Heldt, Lukasz and Hong, Lichan and Tay, Yi and Tran, Vinh and Samost, Jonah and Kula, Maciej and Chi, Ed and Sathiamoorthy, Maheswaran},
  title={Recommender Systems with Generative Retrieval},
  booktitle={Advances in Neural Information Processing Systems},
  year={2023},
  pages={10299--10315},
  volume={36},
  publisher={Curran Associates, Inc.},
  doi={10.52202/075280-0452}
}

@inproceedings{wang2024learnable,
  author={Wang, Wenjie and Bao, Honghui and Lin, Xinyu and Zhang, Jizhi and Li, Yongqi and Feng, Fuli and Ng, See-Kiong and Chua, Tat-Seng},
  title={Learnable Item Tokenization for Generative Recommendation},
  booktitle={Proceedings of the 33rd ACM International Conference on Information and Knowledge Management},
  year={2024},
  pages={2400--2409},
  numpages={10},
  series={CIKM '24},
  publisher={Association for Computing Machinery},
  address={New York, NY, USA},
  location={Boise, ID, USA},
  doi={10.1145/3627673.3679569}
}

@article{tang2025think,
  author={Tang, Jiakai and Dai, Sunhao and Shi, Teng and Xu, Jun and Chen, Xu and Chen, Wen and Wu, Jian and Jiang, Yuning},
  title={Think Before Recommend: Unleashing the Latent Reasoning Power for Sequential Recommendation},
  journal={IEEE Transactions on Knowledge and Data Engineering},
  year={2026},
  volume={38},
  number={9},
  pages={6226--6239},
  doi={10.1109/TKDE.2026.3694421}
}

@article{fang2025reason4rec,
  author={Fang, Yi and Wang, Wenjie and Zhang, Yang and Zhu, Fengbin and Wang, Qifan and Feng, Fuli and He, Xiangnan},
  title={{Reason4Rec}: Deliberative User Preference Alignment of Large Language Models for Recommendation},
  journal={IEEE Transactions on Knowledge and Data Engineering},
  year={2026},
  volume={38},
  number={8},
  pages={5352--5365},
  doi={10.1109/TKDE.2026.3699701}
}

@inproceedings{bismay2024reasoningrec,
  author={Bismay, Millennium and Dong, Xiangjue and Caverlee, James},
  title={{ReasoningRec}: Bridging Personalized Recommendations and Human-Interpretable Explanations through {LLM} Reasoning},
  booktitle={Findings of the Association for Computational Linguistics: NAACL 2025},
  year={2025},
  pages={8147--8163},
  publisher={Association for Computational Linguistics},
  address={Albuquerque, New Mexico},
  doi={10.18653/v1/2025.findings-naacl.454}
}

@inproceedings{gu2025r,
  author={Gu, Hao and Zhong, Rui and Xia, Yu and Yang, Wei and Lu, Chi and Jiang, Peng and Gai, Kun},
  title={{R$^4$ec}: A Reasoning, Reflection, and Refinement Framework for Recommendation Systems},
  booktitle={Proceedings of the Nineteenth ACM Conference on Recommender Systems},
  year={2025},
  pages={411--421},
  numpages={11},
  series={RecSys '25},
  publisher={Association for Computing Machinery},
  address={New York, NY, USA},
  doi={10.1145/3705328.3748068}
}

@inproceedings{wei2022chain,
  author={Wei, Jason and Wang, Xuezhi and Schuurmans, Dale and Bosma, Maarten and Ichter, Brian and Xia, Fei and Chi, Ed H. and Le, Quoc V. and Zhou, Denny},
  title={Chain-of-Thought Prompting Elicits Reasoning in Large Language Models},
  booktitle={Advances in Neural Information Processing Systems},
  year={2022},
  pages={24824--24837},
  volume={35},
  publisher={Curran Associates, Inc.},
  doi={10.52202/068431-1800}
}

@misc{achiam2023gpt,
  author={Achiam, Josh and Adler, Steven and Agarwal, Sandhini and Ahmad, Lama and Akkaya, Ilge and Aleman, Florencia Leoni and Almeida, Diogo and Altenschmidt, Janko and Altman, Sam and Anadkat, Shyamal and others},
  title={{GPT-4} Technical Report},
  year={2024},
  eprint={2303.08774},
  archivePrefix={arXiv},
  primaryClass={cs.CL}
}

@misc{liu2024deepseek,
  author={Liu, Aixin and Feng, Bei and Xue, Bing and Wang, Bingxuan and Wu, Bochao and Lu, Chengda and Zhao, Chenggang and Deng, Chengqi and Zhang, Chenyu and Ruan, Chong and others},
  title={{DeepSeek-V3} Technical Report},
  year={2025},
  eprint={2412.19437},
  archivePrefix={arXiv},
  primaryClass={cs.CL}
}

@inproceedings{su2025token,
  author={Su, Dijia and Zhu, Hanlin and Xu, Yingchen and Jiao, Jiantao and Tian, Yuandong and Zheng, Qinqing},
  title={Token Assorted: Mixing Latent and Text Tokens for Improved Language Model Reasoning},
  booktitle={Proceedings of the 42nd International Conference on Machine Learning},
  year={2025},
  pages={57144--57163},
  series={Proceedings of Machine Learning Research},
  volume={267},
  publisher={PMLR}
}

@inproceedings{hao2024training,
  author={Hao, Shibo and Sukhbaatar, Sainbayar and Su, DiJia and Li, Xian and Hu, Zhiting and Weston, Jason E and Tian, Yuandong},
  title={Training Large Language Models to Reason in a Continuous Latent Space},
  booktitle={Second Conference on Language Modeling},
  year={2025},
  url={https://openreview.net/forum?id=Itxz7S4Ip3}
}

@inproceedings{yu2025thinkrec,
  author={Yu, Qihang and Fu, Kairui and Lv, Zheqi and Zhang, Shengyu and Wu, Xinhui and Lin, Chen and Wei, Feng and Zheng, Bo and Wu, Fei},
  title={{ThinkRec}: Thinking-based recommendation via {LLM}},
  booktitle={Proceedings of the ACM Web Conference 2026},
  year={2026},
  pages={5698--5709},
  numpages={12},
  series={WWW '26},
  publisher={Association for Computing Machinery},
  address={New York, NY, USA},
  location={United Arab Emirates},
  doi={10.1145/3774904.3792070}
}

@misc{liu2025lares,
  author={Liu, Enze and Zheng, Bowen and Wang, Xiaolei and Zhao, Wayne Xin and Wang, Jinpeng and Chen, Sheng and Wen, Ji-Rong},
  title={{LARES}: Latent Reasoning for Sequential Recommendation},
  year={2025},
  eprint={2505.16865},
  archivePrefix={arXiv},
  primaryClass={cs.IR}
}

@inproceedings{lian2018xdeepfm,
  author={Lian, Jianxun and Zhou, Xiaohuan and Zhang, Fuzheng and Chen, Zhongxia and Xie, Xing and Sun, Guangzhong},
  title={{xDeepFM}: Combining Explicit and Implicit Feature Interactions for Recommender Systems},
  booktitle={Proceedings of the 24th ACM SIGKDD International Conference on Knowledge Discovery \& Data Mining},
  year={2018},
  pages={1754--1763},
  numpages={10},
  series={KDD '18},
  publisher={Association for Computing Machinery},
  address={New York, NY, USA},
  location={London, United Kingdom},
  doi={10.1145/3219819.3220023}
}

@inproceedings{zhou2018deep,
  author={Zhou, Guorui and Zhu, Xiaoqiang and Song, Chenru and Fan, Ying and Zhu, Han and Ma, Xiao and Yan, Yanghui and Jin, Junqi and Li, Han and Gai, Kun},
  title={Deep Interest Network for Click-Through Rate Prediction},
  booktitle={Proceedings of the 24th ACM SIGKDD International Conference on Knowledge Discovery \& Data Mining},
  year={2018},
  pages={1059--1068},
  numpages={10},
  series={KDD '18},
  publisher={Association for Computing Machinery},
  address={New York, NY, USA},
  location={London, United Kingdom},
  doi={10.1145/3219819.3219823}
}

@inproceedings{song2026mtfm,
  author={Song, Xin and Guan, Zhilin and Han, Ruidong and Tang, Binghao and Chen, Tianwen and Li, Bing and Li, Zihao and Zhang, Han and Jiang, Fei and Wang, Qing and Xu, Zikang and Li, Fengyi and Jing, Chunzhen and Yu, Lei and Lin, Wei},
  title={{MTFM}: A Scalable and Alignment-free Foundation Model for Industrial Recommendation in Meituan},
  booktitle={Proceedings of the 32nd ACM SIGKDD Conference on Knowledge Discovery and Data Mining V.2},
  year={2026},
  pages={7978--7988},
  numpages={11},
  series={KDD '26},
  publisher={Association for Computing Machinery},
  address={New York, NY, USA},
  location={Republic of Korea},
  doi={10.1145/3770855.3818452}
}

@inproceedings{hu2022lora,
  author={Hu, Edward J and Shen, Yelong and Wallis, Phillip and Allen-Zhu, Zeyuan and Li, Yuanzhi and Wang, Shean and Wang, Lu and Chen, Weizhu},
  title={{LoRA}: Low-Rank Adaptation of Large Language Models},
  booktitle={International Conference on Learning Representations},
  year={2022},
  url={https://openreview.net/forum?id=nZeVKeeFYf9}
}

@article{harper2015movielens,
  author={Harper, F. Maxwell and Konstan, Joseph A.},
  title={The {MovieLens} Datasets: History and Context},
  journal={ACM Trans. Interact. Intell. Syst.},
  year={2015},
  volume={5},
  number={4},
  articleno={19},
  numpages={19},
  publisher={Association for Computing Machinery},
  doi={10.1145/2827872}
}

@inproceedings{lin2025order,
  author={Lin, Xinyu and Shi, Haihan and Wang, Wenjie and Feng, Fuli and Wang, Qifan and Ng, See-Kiong and Chua, Tat-Seng},
  title={Order-Agnostic Identifier for Large Language Model-based Generative Recommendation},
  booktitle={Proceedings of the 48th International ACM SIGIR Conference on Research and Development in Information Retrieval},
  year={2025},
  pages={1923--1933},
  numpages={11},
  series={SIGIR '25},
  publisher={Association for Computing Machinery},
  address={New York, NY, USA},
  location={Padua, Italy},
  doi={10.1145/3726302.3730053}
}

@article{wang2023causal,
  author={Wang, Wenjie and Lin, Xinyu and Wang, Liuhui and Feng, Fuli and Ma, Yunshan and Chua, Tat-Seng},
  title={Causal Disentangled Recommendation against User Preference Shifts},
  journal={ACM Trans. Inf. Syst.},
  year={2023},
  volume={42},
  number={1},
  articleno={12},
  numpages={27},
  publisher={Association for Computing Machinery},
  doi={10.1145/3593022}
}

@inproceedings{pi2019practice,
  author={Pi, Qi and Bian, Weijie and Zhou, Guorui and Zhu, Xiaoqiang and Gai, Kun},
  title={Practice on Long Sequential User Behavior Modeling for Click-Through Rate Prediction},
  booktitle={Proceedings of the 25th ACM SIGKDD International Conference on Knowledge Discovery \& Data Mining},
  year={2019},
  pages={2671--2679},
  numpages={9},
  series={KDD '19},
  publisher={Association for Computing Machinery},
  address={New York, NY, USA},
  location={Anchorage, AK, USA},
  doi={10.1145/3292500.3330666}
}

@inproceedings{lee2022autoregressive,
  author={Lee, Doyup and Kim, Chiheon and Kim, Saehoon and Cho, Minsu and Han, Wook-Shin},
  title={Autoregressive Image Generation Using Residual Quantization},
  booktitle={Proceedings of the IEEE/CVF Conference on Computer Vision and Pattern Recognition (CVPR)},
  year={2022},
  pages={11523--11532},
  doi={10.1109/CVPR52688.2022.01123}
}

@inproceedings{vaswani2017attention,
  author={Vaswani, Ashish and Shazeer, Noam and Parmar, Niki and Uszkoreit, Jakob and Jones, Llion and Gomez, Aidan N and Kaiser, {\L}ukasz and Polosukhin, Illia},
  title={Attention is All you Need},
  booktitle={Advances in Neural Information Processing Systems},
  year={2017},
  volume={30},
  publisher={Curran Associates, Inc.},
  url={https://proceedings.neurips.cc/paper_files/paper/2017/file/3f5ee243547dee91fbd053c1c4a845aa-Paper.pdf}
}

@misc{bai2023qwen,
  author={Bai, Jinze and Bai, Shuai and Chu, Yunfei and Cui, Zeyu and Dang, Kai and Deng, Xiaodong and Fan, Yang and Ge, Wenbin and Han, Yu and Huang, Fei and others},
  title={{Qwen} Technical Report},
  year={2023},
  eprint={2309.16609},
  archivePrefix={arXiv},
  primaryClass={cs.CL}
}

@misc{touvron2023llama,
  author={Touvron, Hugo and Lavril, Thibaut and Izacard, Gautier and Martinet, Xavier and Lachaux, Marie-Anne and Lacroix, Timoth{\'e}e and Rozi{\`e}re, Baptiste and Goyal, Naman and Hambro, Eric and Azhar, Faisal and others},
  title={{LLaMA}: Open and Efficient Foundation Language Models},
  year={2023},
  eprint={2302.13971},
  archivePrefix={arXiv},
  primaryClass={cs.CL}
}

@inproceedings{singer2022sequential,
  author={Singer, Uriel and Roitman, Haggai and Eshel, Yotam and Nus, Alexander and Guy, Ido and Levi, Or and Hasson, Idan and Kiperwasser, Eliyahu},
  title={Sequential Modeling with Multiple Attributes for Watchlist Recommendation in E-Commerce},
  booktitle={Proceedings of the Fifteenth ACM International Conference on Web Search and Data Mining},
  year={2022},
  pages={937--946},
  numpages={10},
  series={WSDM '22},
  publisher={Association for Computing Machinery},
  address={New York, NY, USA},
  location={Virtual Event, AZ, USA},
  doi={10.1145/3488560.3498453}
}

@inproceedings{dai2024modeling,
  author={Dai, Sunhao and Shao, Ninglu and Zhu, Jieming and Zhang, Xiao and Dong, Zhenhua and Xu, Jun and Dai, Quanyu and Wen, Ji-Rong},
  title={Modeling User Attention in Music Recommendation},
  booktitle={2024 IEEE 40th International Conference on Data Engineering (ICDE)},
  year={2024},
  pages={761--774},
  publisher={IEEE},
  location={Utrecht, Netherlands},
  doi={10.1109/ICDE60146.2024.00064}
}

@inproceedings{lei2021semi,
  author={Lei, Chenyi and Liu, Yong and Zhang, Lingzi and Wang, Guoxin and Tang, Haihong and Li, Houqiang and Miao, Chunyan},
  title={{SEMI}: A Sequential Multi-Modal Information Transfer Network for E-Commerce Micro-Video Recommendations},
  booktitle={Proceedings of the 27th ACM SIGKDD Conference on Knowledge Discovery \& Data Mining},
  year={2021},
  pages={3161--3171},
  numpages={11},
  series={KDD '21},
  publisher={Association for Computing Machinery},
  address={New York, NY, USA},
  location={Virtual Event, Singapore},
  doi={10.1145/3447548.3467189}
}

@inproceedings{kojima2022large,
  author={Kojima, Takeshi and Gu, Shixiang (Shane) and Reid, Machel and Matsuo, Yutaka and Iwasawa, Yusuke},
  title={Large Language Models are Zero-Shot Reasoners},
  booktitle={Advances in Neural Information Processing Systems},
  year={2022},
  pages={22199--22213},
  volume={35},
  publisher={Curran Associates, Inc.},
  doi={10.52202/068431-1613}
}

@inproceedings{guo2025onesug,
  author={Guo, Xian and Chen, Ben and Wang, Siyuan and Yang, Ying and Cheng, Mingyue and Lei, Chenyi and Ding, Yuqing and Li, Han},
  title={{OneSug}: The Unified End-to-End Generative Framework for E-commerce Query Suggestion},
  booktitle={Proceedings of the AAAI Conference on Artificial Intelligence},
  year={2026},
  volume={40},
  pages={14774--14782},
  publisher={AAAI Press},
  address={Washington, DC, USA},
  location={Singapore},
  doi={10.1609/aaai.v40i17.38497}
}

@inproceedings{wang2025mtgrboost,
  author={Wang, Yuxiang and Ma, Chi and Yan, Xiao and Huang, Mincong and Li, Xiaoguang and Han, Ruidong and Yin, Bin and Chen, Shangyu and Li, Xiang and Jiang, Fei and Yu, Lei and Liu, Chuan and Lin, Wei and Han, Haowei and Zhou, Xiaokai and Du, Bo and Jiang, Jiawei},
  title={{MTGenRec}: An Efficient Distributed Training System for Generative Recommendation Models in Meituan},
  booktitle={Proceedings of the 32nd ACM SIGKDD Conference on Knowledge Discovery and Data Mining V.1},
  year={2026},
  pages={2482--2493},
  numpages={12},
  series={KDD '26},
  publisher={Association for Computing Machinery},
  address={New York, NY, USA},
  location={Republic of Korea},
  doi={10.1145/3770854.3783925}
}

@inproceedings{guo2017deepfm,
  author={Guo, Huifeng and Tang, Ruiming and Ye, Yunming and Li, Zhenguo and He, Xiuqiang},
  title={{DeepFM}: A Factorization-Machine based Neural Network for {CTR} Prediction},
  booktitle={Proceedings of the Twenty-Sixth International Joint Conference on Artificial Intelligence, {IJCAI-17}},
  year={2017},
  pages={1725--1731},
  doi={10.24963/ijcai.2017/239}
}

@inproceedings{zhang2025reinforced,
  author={Zhang, Yang and Xu, Wenxin and Zhao, Xiaoyan and Wang, Wenjie and Feng, Fuli and He, Xiangnan and Chua, Tat-Seng},
  title={Reinforced Latent Reasoning for {LLM}-based Recommendation},
  booktitle={The Fourteenth International Conference on Learning Representations},
  year={2026},
  url={https://openreview.net/forum?id=eUtIZT2ONS}
}

@inproceedings{zhang2024saqrec,
  author={Zhang, Kepu and Shi, Teng and Dai, Sunhao and Zhang, Xiao and Li, Yinfeng and Lu, Jing and Zang, Xiaoxue and Song, Yang and Xu, Jun},
  title={{SAQRec}: Aligning Recommender Systems to User Satisfaction via Questionnaire Feedback},
  booktitle={Proceedings of the 33rd ACM International Conference on Information and Knowledge Management},
  year={2024},
  pages={3165--3175},
  numpages={11},
  series={CIKM '24},
  publisher={Association for Computing Machinery},
  address={New York, NY, USA},
  location={Boise, ID, USA},
  doi={10.1145/3627673.3679643}
}

@inproceedings{dai2026onepiece,
  author={Dai, Sunhao and Tang, Jiakai and Wu, Jiahua and Wang, Kun and Zhu, Yuxuan and Chen, Bingjun and Hong, Bangyang and Zhao, Yu and Fu, Cong and Wu, Kangle and Ni, Yabo and Zeng, Anxiang and Wang, Wenjie and Chen, Xu and Xu, Jun and Ng, See-Kiong},
  title={{OnePiece}: Bringing Context Engineering and Reasoning to Industrial Cascade Ranking System},
  booktitle={Proceedings of the 32nd ACM SIGKDD Conference on Knowledge Discovery and Data Mining V.2},
  year={2026},
  pages={7117--7128},
  numpages={12},
  series={KDD '26},
  publisher={Association for Computing Machinery},
  address={New York, NY, USA},
  location={Republic of Korea},
  doi={10.1145/3770855.3818456}
}

@inproceedings{lin2026care,
  author={Lin, Xinyu and Liu, Pengyuan and Wang, Wenjie and Hu, Yicheng and Xu, Chen and Feng, Fuli and Wang, Qifan and Chua, Tat-Seng},
  title={Bringing Reasoning to Generative Recommendation Through the Lens of Cascaded Ranking},
  booktitle={Proceedings of the ACM Web Conference 2026},
  year={2026},
  pages={8939--8949},
  numpages={11},
  series={WWW '26},
  publisher={Association for Computing Machinery},
  address={New York, NY, USA},
  location={United Arab Emirates},
  doi={10.1145/3774904.3792980}
}

@misc{tang2026parallel,
  author={Tang, Jiakai and Chen, Xu and Chen, Wen and Wu, Jian and Jiang, Yuning and Zheng, Bo},
  title={Parallel Latent Reasoning for Sequential Recommendation},
  year={2026},
  eprint={2601.03153},
  archivePrefix={arXiv},
  primaryClass={cs.IR}
}

@inproceedings{yang2026beyond,
  author={Yang, Xiaochen and Xu, Yin and Wang, Xiaobei and Wang, Hao and Jiang, Zhida and Chen, Zhen and Xing, Zhaolong},
  title={Beyond Static Representation: Coarse-to-Fine Dynamic Latent Reasoning for Sequential Recommendation},
  booktitle={Database Systems for Advanced Applications},
  year={2026},
  pages={258--273},
  series={Lecture Notes in Computer Science},
  volume={16535},
  publisher={Springer Nature Singapore},
  address={Singapore},
  doi={10.1007/978-981-92-0363-5_16}
}

@inproceedings{shan2026calr,
  title={{CaLR}: Context-Aware Latent Reasoning for Sequential Recommendation},
  author={Shan, Minxuan and Bi, Shuxian and Zhang, Ziye and Dai, Sunhao and Wang, Chuyuan and Li, Yu and Chen, Xiaoshuang and Liu, Yongqi and Zhan, Kaiqiao and Wang, Wenjie},
  booktitle={Proceedings of the 35th ACM International Conference on Information and Knowledge Management},
  year={2026}
}

\end{document}